\documentclass[aps,prb,twocolumn,superscriptaddress,showpacs,notitlepage]{revtex4-1}
\usepackage{amssymb}
\usepackage{amsmath}
\usepackage{appendix}
\usepackage{bm}
\usepackage{graphicx}
\usepackage{txfonts}
\usepackage{color}
\usepackage{rotating}

\newcommand{\bea}{\begin{eqnarray}}
\newcommand{\eea}{\end{eqnarray}}
\newcommand{\bei}{\begin{itemize}}
\newcommand{\eei}{\end{itemize}}
\newcommand{\be}{\begin{equation}}
\newcommand{\ee}{\end{equation}}
\newcommand{\bse}{\begin{subequations}}
\newcommand{\ese}{\end{subequations}}
\newcommand{\bfg}{\begin{figure}}
\newcommand{\efg}{\end{figure}}

\newcommand{\eins}{\mbox{$1 \hspace{-1.0mm} {\bf l}$}}
\newcommand{\zeins}{\mbox{$\hspace{-1.0mm} {\bf 0}$}}

\newcommand{\smeq}{\! = \!}

\newcommand{\smap}{\! \approx \!}

\newcommand{\smpl}{\! + \!}

\newcommand{\smmi}{\! - \!}

\newcommand{\ve}{\varepsilon}

\newcommand{\ci}{\mathrm{i}}

\newcommand{\braket}[1]{\left<#1\right>}

\newcommand{\ketLR}[1]{\left|#1\right>}
\newcommand{\ket}[1]{\left|#1\right>}

\newcommand{\braLR}[1]{\left<#1\right|}

\newcommand{\bese}{\begin{subequations}}
\newcommand{\eese}{\end{subequations}}

\makeatletter
\def\Ddots{\mathinner{\mkern1mu\raise\p@
\vbox{\kern7\p@\hbox{.}}\mkern2mu
\raise4\p@\hbox{.}\mkern2mu\raise7\p@\hbox{.}\mkern1mu}}
\makeatother

\begin{document}
\title{Complex-band structure eigenvalue method adapted to Floquet systems:\\ topological superconducting wires as a case study}
\author{Andres A. Reynoso}
\affiliation{Centre for Engineered Quantum Systems, School of Physics, The University of Sydney, NSW 2006, Australia}
\author{Diego Frustaglia}
\affiliation{Departamento de F\'isica Aplicada II, Universidad de Sevilla, E-41012 Sevilla, Spain}
\date{\today}
\begin{abstract}
For systems that can be modeled as a single-particle lattice extended along a privileged direction as, e.g., quantum wires, the so-called eigenvalue method provides full information about the propagating and evanescent modes as a function of energy. This complex-band structure method can be applied either to lattices consisting of an infinite succession of interconnected layers described by the same local Hamiltonian or to superlattices: Systems in which the spatial periodicity involves more than one layer. Here, for time-dependent systems subject to a periodic driving, we present an adapted version of the superlattice scheme capable of obtaining the Floquet states and the Floquet quasienergy spectrum. Within this scheme the time periodicity is treated as existing along spatial dimension added to the original system. The solutions at a single energy for the enlarged artificial system provide the solutions of the original Floquet problem. The method is suited for arbitrary periodic excitations including strong and anharmonic drivings. We illustrate the capabilities of the methods for both time-independent and time-dependent systems by discussing: (a) topological superconductors in multimode quantum wires with spin-orbit interaction and (b) microwave driven quantum dot in contact with a topological superconductor.
\end{abstract}
\pacs{71.15.-m, 73.21.Hb, 85.35.Be, 74.78.Na, 71.70.Ej, 05.30.Rt}
\maketitle
\section{Introduction}

Systems and devices in which a privileged spatial direction exists are ubiquitous in nature. Translation invariance or periodicity along a given direction is common to many quantum-coherent systems with potential applications in electronics, spintronics and quantum science as, e.g., quantum wires, graphene nanoribbons, nanotubes and molecules. The single-particle description of the carriers in these Bloch systems has been successful to explain and predict measurements in a great deal of experiments in which interaction effects are not significant.\cite{Dattabook} Besides, a substantial amount of research has been devoted to the study of the time evolution of quantum systems subject to periodical drivings (Floquet systems).\cite{KohlerReview2005} Interestingly, such systems have been recently shown to be potential platforms for topological physics.\cite{Kitagawa2010prb,*Kitagawa2010pra,*KitagawaWhiteQW2011,Oka2009prb,Rivera2009,FoaCalvoPastawski2011,Zhou2011prb,Busl2012,FertigPRL2011,FloquetTopoInsNatPhys2011}

In Floquet systems, the energy $E$ is not conserved. However, solutions of the driven Schroedinger equation can still be classified by resorting to the concept of quasienergy. At a given time $t_0$, a solution with quasienergy $\ve$ reads $\ketLR{\phi_\ve(t_0)}$. After a time equal to the driving period, $T$, such state looks the same up to a phase factor given by the quasienergy: ${\rm exp}(-\ci\ve T/\hbar ) \ketLR{\phi_\ve(t_0)}$. Here, we present a scheme to find the Floquet states and quasienergies which is based on the eigenvalue method [usually applied to calculate the bands of time-independent translationally invariant and periodic lattice systems (Bloch systems)].\cite{Wachutka1986,RamMohan1988,Ando89,Schulman1991} The method was recently applied in Ref.\onlinecite{ReynosoFrustaglia2013a} to the study of Floquet topological transitions. To this aim, the time axis is regarded as an effective spatial dimension which is discretized generating a lattice. For $d$-dimensional systems, the extended lattice is $d+1$-dimensional with a periodicity along the additional dimension. At this point, the system is solved by assuming that it is subject to a particular Schroedinger equation governing the dynamics along an artificial time, $\tilde{t}$. In this new parametrization, a conserved artificial energy $\tilde{E}$ exists. However, only the $\tilde{E}\smeq 0$ solutions are physically meaningful for the original problem as these have no dynamics along the artificial time.\footnote{Due to the discretization procedure along $t$ some non-physical $\tilde{E}\smeq 0$ solutions arise and must be discarded, see Sec.\ref{SC:Floquet}.} The quasienergies of these Floquet solutions are extracted from the pseudomomenta along the spatial dimension associated with the physical time. We establish the conditions on the discretization time step for an accurate description of the Floquet problem.

The paper is organized as follows. In Sec.\ref{SC:methods} we review the eigenvalue method for Bloch systems and introduce our Floquet implementation. In Sec.\ref{SC:Examples} we provide some examples on the production of Majorana fermions in topological superconductors based on quantum wires with spin-orbit coupling (SOC).\cite{SauDasSarma2010,*Alicea2010prb,*OregRefaelvonOppen2010} We start by studying static (i.e., time independent) situations to obtain bands and the effective superconducting gap. In gapped conditions we use the zero energy evanescent states to compute the topological number by counting the Majorana bound states existing at an end of the wire.\cite{Serra2013} This illustrates how one can identify the topological nature of a gapped phase by applying the complex-band structure method at a single energy value: For translational invariant and periodic systems the present scheme provides an alternative to the scattering matrix approach.\cite{Fulga2011prb} With this tools we discuss how the spin-orbit coupling induces transverse-mode mixing affecting the topological transition.\cite{PotterLee2010,*PotterLee2011,Lutchyn2011,Tewari2012,Serra2013} Quantum wires in experiments are multichannel devices subject to SOC mixing\cite{Kouwenhoven2012,RokhinsonMajorana2012,Das2012,Sasaki2011PRL} which leads to extra gapless conditions with consequences on the generation of Majorana fermions. Secondly, we test the method on superlattice potentials by incorporating a periodic profile along the wire. We show that this introduces new gapless conditions at which the topological number changes. Thirdly, demonstrating our scheme to solve Floquet systems, we consider a topological superconductor in contact with a driven quantum dot in Coulomb blockade. As the dot's energy is periodically excited a critical frequency exists at which the condition for generating Floquet Majorana fermions\cite{Cirac2011prl,Liu2012superfluid,KunduFMF2013,LiuBarangerFMF} is less sensitive to the dot's average energy. Finally, in Sec.\ref{SC:Concl}, we conclude and summarize the results.

\section{Adapted superlattice method for solving Floquet systems}
\label{SC:methods}
\subsection{Eigenvalue Method for time-independent systems}
\label{SC:supl}
We start by considering a mesoscopic/nanoscopic system built upon (or modeled by) a sequence of interconnected, identical layers. The \emph{layer} here plays a role similar to the \emph{unit cell} in crystals: A basic building block grouping a finite set of sites from which the system is constructed. Each site in real space may be represented by more than one site in the tight-binding lattice when the spin and/or the electron/hole degrees of freedom are relevant to the problem (for instance, a single spin-1/2 site is represented by \emph{two} effective sites in normal system and by \emph{four} effective sites in a superconducting one). For a layer with $N$ effective lattice sites, the Hamiltonian $\hat{H}_0$ is a $N\times N$ Hermitian matrix. A second ingredient required to define the full Hamiltonian $\hat{H}$ is the hopping matrix connecting first nearest-neighbor layers, $\hat{T}_0$.

The problem reduces to the one of finding the eigenvalues $E$ and eigenvectors $\ket{\Psi}$ in $\hat{H} \ket{\Psi}=E \ket{\Psi}$ (the solution of the original time-dependent Schroedinger equation is $(\hat{H}-\ci \hbar\frac{d}{d t}) \ket{\Psi(t)}$, is $\ket{\Psi(t)}=\ket{\Psi}\mathrm{e}^{-\ci E t/\hbar}$). In the layer-block matrix representation, the equation $\hat{H} \ket{\Psi}=E \ket{\Psi}$ becomes
\begin{equation}
 \left(
\begin{array}{ccccc}  \ddots & \vdots  &  \vdots  & \vdots   &  \Ddots \\ \cdots & \hat{H}_{0}  & \hat{T}_0 & \zeins_{N} & \cdots
\\ \cdots   &  \hat{T}_0^\dagger & \hat{H}_{0} & \hat{T}_0 & \cdots
 \\ \cdots &  \zeins_{N} & \hat{T}_0^\dagger & \hat{H}_{0} &  \cdots \\ \Ddots & \vdots  &  \vdots  & \vdots   &  \ddots
\end{array} \right) \left(
\begin{array}{c} \vdots  \\  G_{0} \\ G_{1} \\ G_{2} \\ \vdots
\end{array} \right) = E \left(
\begin{array}{c} \vdots  \\   G_{0} \\ G_{1} \\ G_{2}  \\ \vdots
\end{array} \right),
\label{EQ:Hfull}
\end{equation}
with $G_{i}=(g_{i,1},g_{i,2}, \cdots,g_{i,N})^T$ a $N$-dimensional vector representing the eigenstate $\ket{\Psi}$ within the layer $i$ (in the basis in which the matrix $\hat{H}_0$ is written) and $~~\zeins_{N}$ the null $N\times N$ square matrix.

For each layer $i$, Eq.~(\ref{EQ:Hfull}) leads to the relation $\hat{T}_0^\dagger G_{i-1} +  \hat{T}_0 G_{i+1}=(E \eins_N-\hat{H}_0)G_{i}$ with $\eins_N$ the $N$-dimensional identity matrix. Notice that if the amplitudes of a solution with energy $E$ at two consecutive layers are known one can obtain the amplitude at the next layer through a transfer-matrix method:
\be
\bar{G}_{i+1}= \mathcal{M}(E)  \bar{G}_{i}
\ee
where $\bar{G}_{i}\equiv(G_{i},G_{i-1})^T$ and the transfer matrix, $\mathcal{M}(E)$, is the $2N\times2N$ matrix
\be
\mathcal{M}(E)\equiv \left(\begin{array}{cc}
 \hat{T}_0^{-1}\left(E\eins_N -\hat{H}_0 \right)& -\hat{T}_0^{-1}\hat{T}_0^\dagger \\
\eins_N & \zeins_N
\end{array}\right).
\label{EQ:TransMat}
\ee

\begin{figure}[t]\begin{center}
\includegraphics[width=0.32\textwidth]{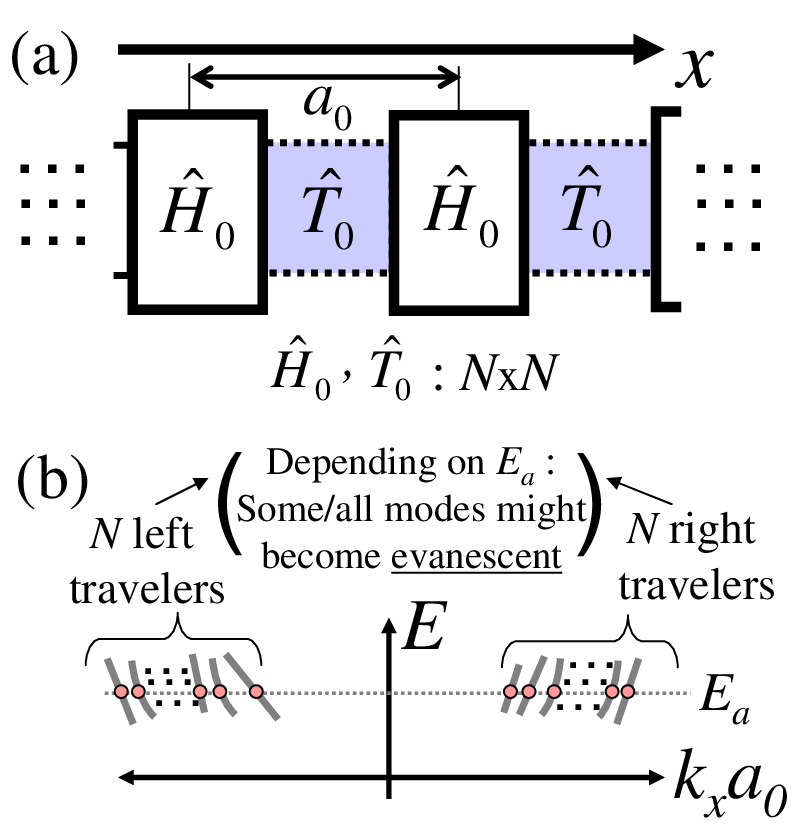}
\vspace{-0.1cm}
\caption{\label{FG:fig1} (a) Translation invariant single-layer periodical lattice system: Each layer has $N$ effective sites described by the $N\times N$ Hamiltonian $\hat{H}_0$, while the $N\times N$ hopping matrix $\hat{T}_0$ describes the hopping between neighbor layers, which are separated by a distance $a_0$. (b) The eigenvalue method takes $\hat{H}_0$ and $\hat{T}_0$ as the input, once an energy value $E_a$ is chosen it allows---provided $\hat{T}_0^{-1}$ exists (see text)---for the obtention of the allowed values of the wavevector $k_x$ and the corresponding eigenstates. For energies in which all quantum channels are open there are $N$ left-traveling and $N$ right-traveling states. In general, at other energies, some (or all) of the solutions become evanescent: $\mathrm{Im}[k_x]\neq 0$. As sketched in the figure, at each simulated energy $E_a$ one finds solutions (points) which pertain to bands (lines) and therefore, by sweeping $E_a$ the underlying full dispersion relation, $E(k_x)$, can be obtained.}
\end{center}
\vspace{-0.5cm}
\end{figure}

Furthermore, the different $G_{i}$ are related by virtue of the \emph{single-layer} periodicity shown in Fig.\ref{FG:fig1}(a). This series of layers defines a direction, here denoted by $x$, along which the system has discrete translation invariance leading to the conservation of a wavenumber, $k_x$. Assuming the distance between layers is $a_0$ then $\hat{H}(x)=\hat{H}(x+a_0)$ and Bloch theorem can be applied. Since each layer (or unit cell) has $N$ effective sites, the solutions can be classified as $\ket{\Psi_{n,k_x}}$ with energies $E_n(k_x)$ and $n=\{1,2,..,N\}$. The Bloch theorem states that
\be
\Psi_{n,k_x} (x) ={\rm e }^{\ci k_x x} \psi_{n,k_x} (x),
\label{EQ:Bloch}
\ee
with $\psi_{n,k_x} (x)=\psi_{n,k_x} (x+a_0)$. This implies that, in terms of the $G_i$ vectors, a solution for a given $k_x$ at a given \emph{band} $n$ must fulfill
\be
G_i= {\rm e }^{\ci k_x a_0} G_{i-1} = \lambda G_{i-1},
\ee
with $\lambda\equiv {\rm e }^{\ci k_x a_0}$.
\begin{figure}[!b]
\includegraphics[width=0.45\textwidth]{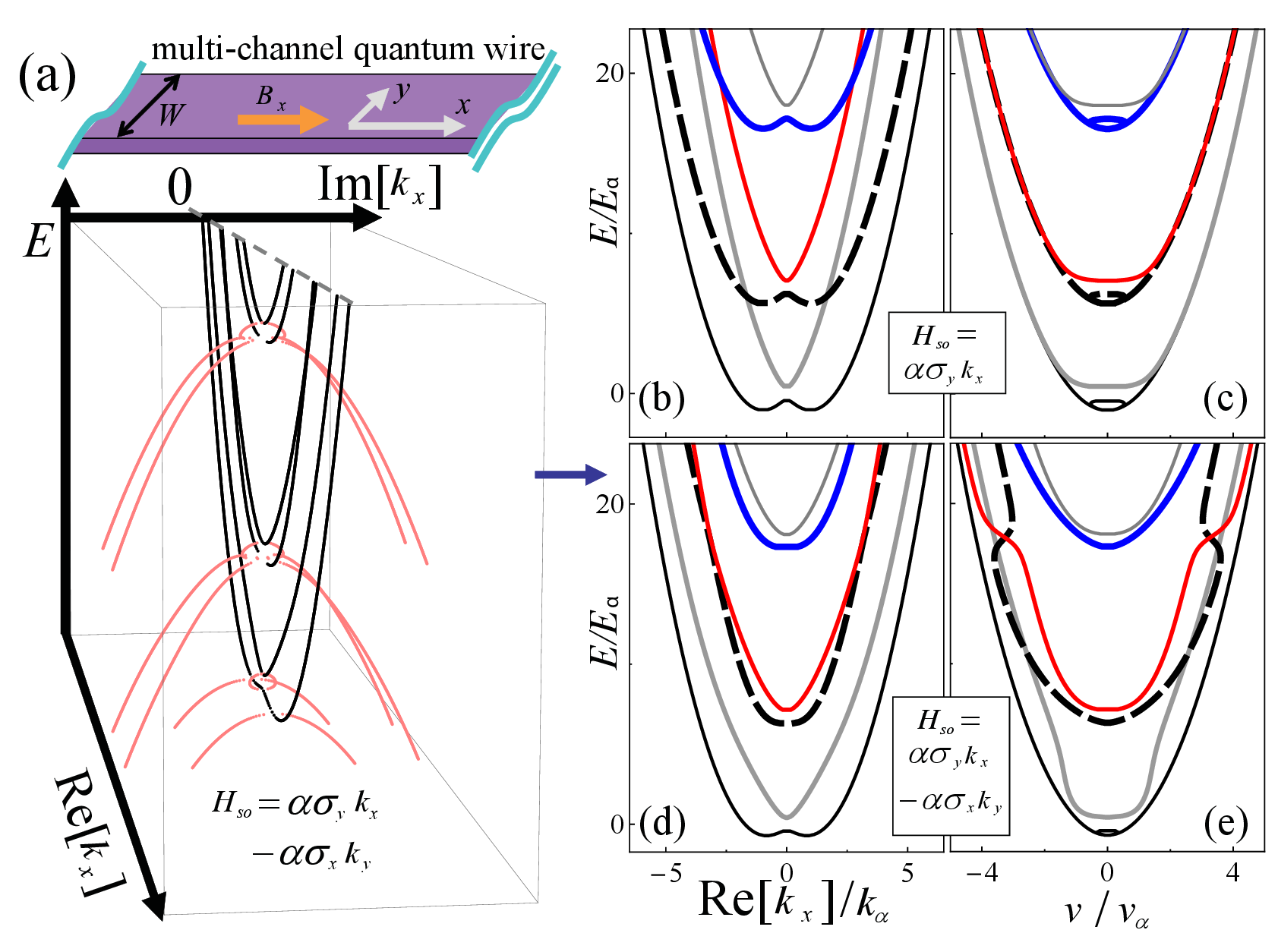}
\vspace{-0.1cm}
\caption{\label{FG:fig2} (color online) Complex band structure, dispersion relations and velocities of the traveling states as obtained with the method for the case of a multichannel quantum wire described in Sec.\ref{SC:EgQWire}. (a) The sketch shows the sample subject to an inplane magnetic field $B_x$; the longitudinal SOC term, $\alpha_\parallel$, is fixed at a nonzero value (not shown) and we use as reference energy $E_\alpha\equiv \alpha_\parallel^2 m^*/2\hbar^2  \smeq \hbar^2 k_\alpha^2/2 m^*$, reference momentum $k_\alpha\smeq \alpha_\parallel m^*/\hbar^2$ and reference velocity $v_\alpha\smeq\alpha_\parallel/\hbar$. Parameters are: Zeeman energy $E_Z\equiv g\mu_B B_x /2 = 0.4 E_\alpha$, sample width $W$ is related to the SOC-length as $L_\alpha\smeq 2\pi k_\alpha^{-1}  \smeq \frac{20}{21}\pi W \approx  3 W$ (i.e., the $n$-transverse mode energy at zero field for $k_x\smeq0$ is $E_n\smap \frac{9}{4} E_\alpha n^2 \smpl E_\alpha \alpha_\perp^2/\alpha_\parallel^2$). The 3D plot shows a typical complex band structure: evenescent states (pink line) have nonzero $\mathrm{Im}[k_x]$, the relation dispersion (black line) is contained in the plane $\mathrm{Re}[k_x]-E$. From the full complex band structure we group the traveling states in bands to obtain the dispersion relations for $\alpha_\perp\smeq 0$ [in (b)] and for $\alpha_\perp\smeq \alpha_\parallel$ [in (d)], $E_1$ has been subtracted to align $E=0$ with the $k_x\smeq 0$ degeneracy point of the first transverse mode when $B_x\smeq0$. The perpendicular component of the SOC in the Hamiltonian, $\alpha_\perp k_y \sigma_x$, generates mixing between transverse modes with different spin and parity. The velocity $v$ is computed for each energy $E$ traveling state and plotted in panels (c) and (e): the effect of nonzero $\alpha_\perp$ in (e) is clearly observed.}
\vspace{-0.1cm}
\end{figure}

The eigenvalue method is derived when combining the Bloch theorem and Eq.\eqref{EQ:TransMat}, finding $\mathcal{M}(E)\bar{G}_{i}=\lambda \bar{G}_{i}$.\cite{Wachutka1986,RamMohan1988,Ando89,Schulman1991} By fixing the energy at a given value, $E_a$ (see Fig.\ref{FG:fig1}(b)), we obtain
\be
\mathcal{M}(E_a) \bar{G}^{(l)}_{i}=\lambda_l \bar{G}^{(l)}_{i} ,
\ee
with $l=1,2,..,2N$ and $\bar{G}^{(l)}_{i}$ an eigenvector of $\mathcal{M}(E_a)$ with eigenvalue $\lambda_l$. By sweeping $E_a$ within a range of interest, the dispersion relations $E_n(k_x)$ are found. For traveling solutions, it holds  $|\lambda_l|=1$ and the wavenumber is found from
\be
k_x= \frac{1}{a_0} \mathrm{Arg} \lambda_l.
\ee
In the general case there will be a nonzero number of $\mathcal{M}(E_a)$ eigenvalues with $|\lambda_l|\neq 1$.  These correspond to evanescent solutions which have complex wavenumbers
\be
k_x= \frac{1}{a_0} \left(\mathrm{Arg} \lambda_l - \ci \ln|\lambda_l| \right) .
\label{EQ:complexKX}
\ee
In all cases the $2N$-dimensional vector $\bar{G}^{(l)}_{i}$ has $N$ redundant components as $\bar{G}^{(l)}_{i}=(\lambda_l G^{(l)}_{i-1},G^{(l)}_{i-1})^T$ by virtue of the Bloch theorem in the form $G^{(l)}_{i}=\lambda_l G^{(l)}_{i-1}$.

The velocity, $v_l$, of a given traveling solution can be investigated numerically by simulating two very close energies $E_a$ and $E_a+\Delta E$ in order to estimate the derivative $\hbar^{-1} \partial E_n/\partial k_x$: The energy step must be sufficiently small to assure a small change in the phase of $\lambda_l$ and a large overlap between the eigenvectors $G^{(l)}_{i}$ at the two energies. A convenient alternative is to use  the velocity operator $\dot{\hat{X}}=\ci[\hat{H},\hat{X}]/\hbar$: For a solution $\ket{\Psi}$ the velocity is obtained as $v=\braket{\Psi|\dot{\hat{X}}|\Psi}$, this quantity depends on the amplitudes of $\ket{\Psi}$ at any two neighbor layers $i$ and $i+1$.\cite{Bruus2004book} For each traveling solution the latter amplitudes can be readily obtained from $G^{(l)}_{i}$ and $\lambda_l$; the velocity of the state is given by
\bea
\hbar v_l&=&  \ci a_0 \left[\lambda_l \left(G^{(l)}_{i} \right)^\dagger \hat{T}_0  G^{(l)}_{i} - \lambda_l^* \left(G^{(l)}_{i} \right)^\dagger \hat{T}_0^\dagger  G^{(l)}_{i}\right] \notag \\
&=& -2 a_0 \mathrm{Im}\left\{\lambda_l \left(G^{(l)}_{i} \right)^\dagger \hat{T}_0  G^{(l)}_{i}\right\}.
\label{EQ:velocity}
\eea
Notice that $\left(G^{(l)}_{i} \right)^\dagger \hat{T}_0  G^{(l)}_{i}$ is a complex number, which results from the summation $\sum_{r=1}^N \sum_{s=1}^N  \left(g^{(l)}_{i,r} \right)^* \left[\hat{T}_{0}\right]_{r,s} g^{(l)}_{i,s}$.

In Fig.\ref{FG:fig2} we present some typical results obtained with this method for the case of a quantum wire. Figure \ref{FG:fig2}(a) shows a full complex band structure. Figure \ref{FG:fig2}(b) shows the dispersion relations: $E(k_x)$ for the traveling solutions. And Fig.\ref{FG:fig2}(c) shows the velocities associated with the traveling eigenstates. Details on $\hat{T}_0$ and $\hat{H}_0$ for this quantum wire, which is subject to spin-orbit and Zeeman couplings, are given in Sec.\ref{SC:EgQWire}.

\subsubsection*{Superlattices}

In the case of superlattices, each block or unit cell involves $P>1$ layers: the set of $P$ consecutive layers forms a \emph{superlayer}. Each layer $i$ may have different number of effective sites, $N_i$. The full Hamiltonian is given by the layer Hamiltonians $\{\hat{H}_i | i={1,2,\cdots,P}\}$ and by the hopping operators connecting neighboring layers. [Within a superlayer, these are $\{\hat{T}_{1,2},\hat{T}_{2,3},..,\hat{T}_{P-1,P}\}$. An additional $\hat{T}_{P,1}$ connects the last layer of a superlayer with the first layer of the next superlayer]. Notice that the hopping operators $\hat{T}_{i,j}$ ($\hat{T}_{i,j}^\dagger$) are represented by $N_i\times N_j$ ($N_j\times N_i$) rectangular matrices. One would be tempted to define a superlayer as a single layer with $\tilde{N}=\sum_{i=1}^P N_i$ effective sites in order to build a $2\tilde{N}$-dimensional transfer matrix as in Eq.\eqref{EQ:TransMat}. However, this cannot be done: The hopping matrix between the extreme layers in two superlayers, $\hat{T}_{P,1}$, is at most $\max\{N_1,N_P\}$-dimensional and, therefore, the associated $\tilde{N}$-dimensional superlayer-hopping matrix that contains $\hat{T}_{P,1}$ is not invertible.

\begin{figure}[!t]\begin{center}
\includegraphics[width=0.42\textwidth]{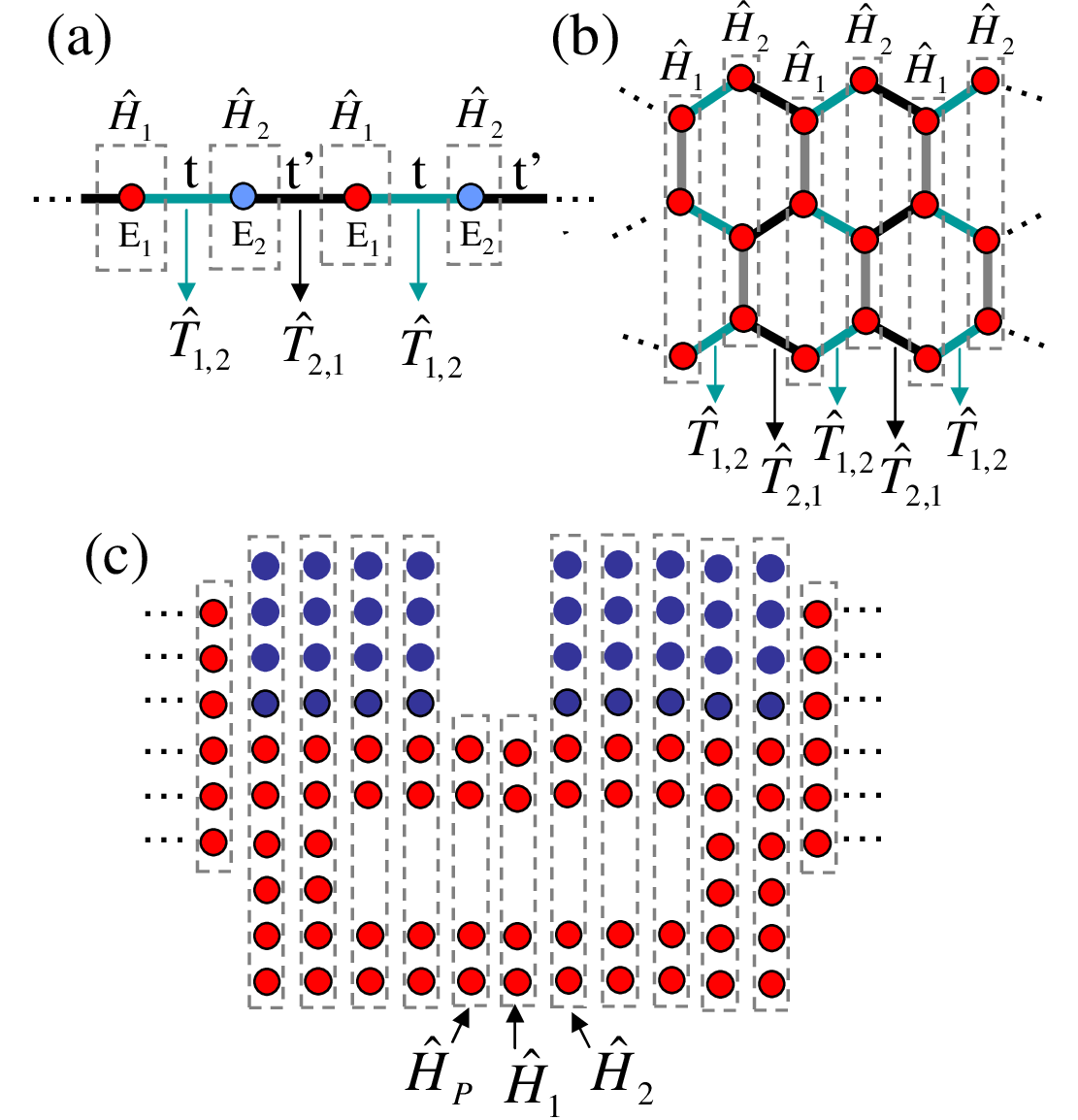}
\vspace{-0.1cm}
\caption{\label{FG:fig3} Examples of superlattices: (a) Tight-binding model with two inequivalent sites having sites energies $E_1$ and $E_2$ and connecting hopping energies $t$ and $t'$, this is trivially casted to a superlattice with $P=2$. (b) A zigzag graphene nanoribbon is also casted to a superlattice with $P=2$. (c) The scheme is valid for superlayers containing layers with different dimensions, inequivalent sites and vacancies (for simplicity the intralayer and interlayer hoppings elements are not shown). By choosing the $i=1$ layer one having the fewest number of effective sites the characteristic matrix dimension.}
\end{center}
\vspace{-0.1cm}
\end{figure}

For arbitrary potentials inside a superlayer, the Schroedinger equation $\hat{H} \ket{\Psi}=E \ket{\Psi}$ would produce different equations for each layer  $i=\{1,2,...,P\}$:
\be
\hat{T}^\dagger_{i-1,i} G_{i-1} +  \hat{T}_{i,i+1} G_{i+1}=(E \eins_{N_i}-\hat{H}_i)G_{i}.
\label{EQ:hamil}
\ee
For illustration, in Fig.\ref{FG:fig3} we sketch some superlattice examples. In the following, we discuss two alternative approaches to calculate the dispersion relation in superlattices. The first approach is based on the transfer matrix scheme, applicable when all layers have the same dimension $N_i$ [see Fig.\ref{FG:fig3}(a) and (b)]. The second one can be applied to the general case [see Fig.\ref{FG:fig3}(c)].

Consider the joint wavefunction amplitude of two neighboring layers $i-1$ and $i$, i.e., the $N_i+N_{i+1}$ dimensional vector $\bar{G}_{i}\equiv(G_{i},G_{i-1})^T$. For identically dimensional layers ($N_i=N$), the
$\bar{G}_{i}$ results $2N$-dimensional. In this case, we can build transfer matrices $\mathcal{M}_i(E)$ inside
a superlayer as
\be
\mathcal{M}_i(E)\equiv \left(\begin{array}{cc}
 \hat{T}_{i,i+1}^{-1}\left(E\eins_N -\hat{H}_i \right)& -\hat{T}_{i,i+1}^{-1}\hat{T}_{i-1,i}^\dagger \\
\eins_N & \zeins_N
\end{array}\right)
\label{EQ:TransMati}
\ee
such that $\bar{G}_{i+1}= \mathcal{M}_i(E)  \bar{G}_{i}$ [one must use $\hat{T}_{P,1}$ to replace $\hat{T}_{i-1,i}$ ($\hat{T}_{i,i+1}$) for the case of $i=1$ ($i=P$)]. For the superlattice Bloch theorem also holds, as the periodic length is $P a_0$ we have that $\bar{G}_{P+i}=\widetilde{\lambda} \bar{G}_{i}$ with $\widetilde{\lambda}\equiv {\rm e }^{\ci  k_x P a_0}$. Therefore the eigenvalue problem can be formulated as
\bea
\bar{G}_{P+1}= \widetilde{\mathcal{M}}(E) \bar{G}_{1} = \widetilde{\lambda}  \bar{G}_{1}
\eea
with
\be
\widetilde{\mathcal{M}}(E)\equiv \mathcal{M}_P(E) \mathcal{M}_{P-1}(E)\cdots \mathcal{M}_2(E) \mathcal{M}_{1}(E).
\ee
The corresponding dispersion relations can be obtained by sweeping $E$ in the range of interest and computing the eigenvalues of $\widetilde{\mathcal{M}}(E)$, $\widetilde{\lambda}_l$. Notice that the eigenstate $\bar{G}^{(l)}_{1}$ only has information about the amplitude of the solution at layers $0$ and $1$. Information on the eigenstate structure within a superlayer can be obtained by propagating the solution layer by layer, using the single-layer transfer matrices $\mathcal{M}_i(E)$. Additionally, the velocity of a traveling solution $\ketLR{\Psi_l}$---the wavefunction associated with $\bar{G}^{(l)}_{1}$---is given as
\be
v_l= \braket{\Psi_l|\dot{\hat{X}}|\Psi_l}= -2 a_0 \hbar^{-1} \mathrm{Im}\left\{\left(G^{(l)}_{0} \right)^\dagger \hat{T}_{P,1}  G^{(l)}_{1}\right\}.
\label{EQ:velocitySuperlattice}
\ee

In this first scheme, the transfer matrix $\widetilde{\mathcal{M}}(E)$ links the wavefunction amplitudes at layers $P+1$ and $P$ to those at layers $1$ and $0$. The second scheme handles different layer dimensions. It is based on a recursive procedure to extract layers out from the superlayer, renormalizing the hopping matrices and the Hamiltonian of the remaining layers. This is repeated $P-1$ times so that one single effective layer is eventually obtained. The procedure starts from Eq.\eqref{EQ:hamil} for three neighboring layers, by discarding the central one: For layers $i=1,2$ and $3$ we obtain $G_{2}$ as a function of $G_1$ and $G_3$ from the equation corresponding to $i=2$. This is introduced into the equations for $i=1$ and $i=3$. In the next iteration one adds the equation for layer $i=4$ to the updated equations for $i=1$ and $i=3$. After $q$ iterations, one finds the wavefunction amplitude at layer $1$ linked to those at layer $q+2$ (where the amplitudes corresponding to all intermediate layers have been iterated out):
\bese
\bea
\hat{T}^\dagger_{P,1} G_{0} +  \hat{T}_f^{(q)} G_{q+2}&=&(E \eins_{N_1}-\hat{H_1}^{(q)})G_{1}, \\
\hat{T}_b^{(q)} G_{1} +  \hat{T}_{q+2,q+3} G_{q+3}&=&(E \eins_{N_{q+2}}-\hat{H}_{q+2}^{(q)})G_{q+2}
\eea
\label{EQ:hamil2}
\eese
Initially, at iteration $q=0$ we have
\be
\hat{T}_f^{(0)}=\hat{T}_{1,2} ~,~~\hat{T}_b^{(0)}=\hat{T}^\dagger_{1,2}~,~~\hat{H}_1^{(0)}=\hat{H}_1~,~~\hat{H}_{2}^{(0)}=\hat{H}_2.
\label{EQ:SupLatAlg0}
\ee
The updating processes at iteration $q$ are
\bese
\bea
 \hat{T}_f^{(q-1)} \left(E \eins_{N_{q+1}}-\hat{H}^{(q-1)}_{q+1}\right)^{-1} \hat{T}_{q+1,q+2}  &\rightarrow& \hat{T}_f^{(q)} \\
 \hat{T}_{q+1,q+2}^\dagger  \left(E \eins_{N_{q+1}}-\hat{H}^{(q-1)}_{q+1}\right)^{-1} \hat{T}_b^{(q-1)} &\rightarrow& \hat{T}_b^{(q)}\\
 \hat{H}_1^{(q-1)} + \hat{T}_f^{(q-1)} \left(E \eins_{N_{q+1}}-\hat{H}^{(q-1)}_{q+1}\right)^{-1} \hat{T}_b^{(q-1)}
&\rightarrow& \hat{H}_1^{(q)} \\
 \hat{H}_{q+2} +  \hat{T}_{q+1,q+2}^\dagger \left(E \eins_{N_{q+1}}-\hat{H}^{(q-1)}_{q+1}\right)^{-1}  \hat{T}_{q+1,q+2} &\rightarrow& \hat{H}_{q+2}^{(q)}~,~~~~
\eea
\label{EQ:SupLatAlg1}
\eese
which are valid for $q<P-1$. For the last iteration ($q=P-1$), a special updating applies since simultaneous operations are taking places at neighboring superlayers: updates of $\hat{H}_{P+1}^{(P-1)}$ and $\hat{H}_1^{(P-1)}$ must merge through the hopping matrix $\hat{T}_{P,1}$ as:
\bea
\hat{H}_1^{(P-2)} + \hat{T}_f^{(P-2)} \left(E \eins_{N_P}-\hat{H}^{(P-2)}_{P}\right)^{-1} \hat{T}_b^{(P-2)}\notag \\ + \hat{T}_{P,1}^\dagger \left(E \eins_{N_P}-\hat{H}^{(P-2)}_{P}\right)^{-1} \hat{T}_{P,1} \rightarrow \hat{H}_1^{(P-1)}.
\label{EQ:SupLatAlg2}
\eea
The updates for $\hat{T}_b^{(P-1)}$ and $\hat{T}_f^{(P-1)}$ remain unaffected.

At the end of the procedure, each superlayer is represented by a single layer of dimension $N_1$. [As an example, in Fig.\ref{FG:fig3}(c), layer 1 is chosen to be the one with the least amount of effective sites.] The distance between effective layers belonging to neighboring superlayers is $P a_0$, so that $\widetilde{\lambda}= {\rm e }^{\ci  k_x P a_0}$. With the help of Eq.\eqref{EQ:TransMat},  the eigenvalue method can be then formulated as
\be
\left(\begin{array}{c}
G_{P+1}\\
G_1\end{array}\right)=\mathcal{M}(E)\left(\begin{array}{c}
G_{1}\\
G_{1-P}\end{array}\right)=\widetilde{\lambda}\left(\begin{array}{c}
G_{1}\\
G_{1-P}\end{array}\right),
\label{EQ:SupLatM}
\ee
after identifying $\hat{H}_0\leftrightarrow \hat{H}_1^{(P-2)}$,  $\hat{T}_0\leftrightarrow \hat{T}_f^{(P-2)}$ and  $\hat{T}^\dagger_0 \leftrightarrow \hat{T}_b^{(P-2)}$. The superlattice problem is in this way casted to a single-layer periodic system. However, it is not strictly equivalent to it since for each energy $E$ it leads to a different set of operators $\hat{H}_0$, $\hat{T}_0$ and $\hat{T}^\dagger_0$. This scheme is well suited for finding the dispersion relation and the wavefunction amplitude at single layers (here layers $i=1+ n P$). The velocity of a traveling solution is obtained from the eigenvectors exactly as in Eq.\eqref{EQ:velocity} by replacing $a_0$ with $P a_0$.

\subsection{Extension to Floquet Systems}
\label{SC:Floquet}
In this section we present a method that allows for the direct application of the calculation scheme of Sec.\ref{SC:supl} to solve Floquet problems. To distinguish the present method from customary approaches we start by revisiting the basic concepts of Floquet theory. Given a time-dependent periodic quantum system, the goal to solve the Schroedinger equation
\be
\left(\hat{H}(t)-\ci \hbar\frac{d}{d t}\right)\ket{\Psi_a(t)}=0,
\label{EQ:Htime}
\ee
where $\hat{H}(t)\smeq\hat{H}(t+T)$ (with $T\smeq \frac{2\pi}{\omega}$ the driving period) and the subindex $a$ labels the quantum numbers of the different solutions. The Floquet theorem states that the solutions for a periodically driven system can be classified by the quasienergy $\ve_a$ as
\be
\ketLR{\Psi_a (t)} = {\rm e}^{-\frac{\ci}{\hbar} \varepsilon_{a} t}\ketLR{\phi^T_{a}(t)},
\label{EQ:physical}
\ee
where $\ketLR{\phi^T_{a}(t)}=\ketLR{\phi^T_{a}(t\smpl T)}$ are the quasienergy states (QESs). Clearly, see Eq.\eqref{EQ:Bloch}, the quasienergy plays (in time) a role similar to the one played (in space) by the wavenumber $k_x$ in spatially periodic systems to which Bloch theorem applies.

In order to obtain the Floquet solutions one substitutes Eq.\eqref{EQ:physical} into the Schroedinger equation \eqref{EQ:Htime} to obtain
\be
\left(\hat{H}(t)-\ci\hbar \frac{d}{dt}\right) \ketLR{\phi^T_{a}(t)}=\varepsilon_{a}\ketLR{\phi^T_{a}(t)}.
\label{EQ:floquetQES}
\ee
Therefore, obtaining the full Floquet spectrum involves finding the set of solutions generated from the existing eigenvalues of the operator $H_{\rm F}\equiv \left(\hat{H}(t)-\ci\hbar \frac{d}{dt}\right)$, provided the latter is restricted to have time-periodic eigenstates.\footnote{The quasienergies and QESs can also be extracted from the evolution operator $\hat{\mathcal{U}}\left(t,t_0\right)\smeq {\bm{T}}_t\exp{\left(-\frac{\ci}{\hbar} \! \int_{t_0}^t  \hat{H}(t') dt' \right)}$, where ${\bm{T}}_t$ stands for time ordering. From Eq.\eqref{EQ:floquetQES} it then follows
$\hat{\mathcal{U}}\left(t_0+T,t_0\right)  \ketLR{\phi^T_{a}(t_0)}=  {\rm e}^{-\frac{\ci}{\hbar}\ve_a T} \ketLR{\phi^T_{a}(t_0)}$. Hence, the quasienergies determine the phase factors ${\rm e}^{-\frac{\ci}{\hbar}\ve_a T}$ which are the eigenvalues of the evolution operator over a driving period.}

Customary approaches make use of the Fourier representation of the Hamiltonian and the Floquet QESs\cite{Sambe1973}
\be
\hat{H}(t)=\sum_{n=-\infty}^{\infty} {\rm e}^{-\ci \omega n t} \hat{H}^{(n)}~,~~\ketLR{\phi^T_{a}(t)}=\sum_{n=-\infty}^{\infty} {\rm e}^{-\ci \omega n t} \ketLR{\phi^{(n)}_{a}}.
\label{EQ:HtFourier}
\ee
By substituting Eq.\eqref{EQ:HtFourier} into Eq.~(\ref{EQ:floquetQES}) one arrives to
$\sum_{m}\left(\hat{H}^{(m)}- n \hbar \omega \delta_{0,m}\right)\ketLR{\phi^{(n-m)}_{a}} = \varepsilon_a  \ketLR{\phi^{(n)}_{a}}$, a time-independent infinite dimensional eigenvalue problem for $\ve_a$ and $\ketLR{\phi^{(n)}_{a}}$. Its matrix representation reads
\begin{widetext}
\begin{equation}
 \left(
\begin{array}{ccccc}  \ddots & \vdots  &  \vdots  & \vdots   &  \Ddots \\ \cdots & \left(\hat{H}^{(0)} + \hbar\omega\right) & \hat{H}^{(-1)} &\hat{H}^{(-2)}& \cdots
\\ \cdots   &  \hat{H}^{(1)} & \hat{H}^{(0)} & \hat{H}^{(-1)} & \cdots
 \\ \cdots & \hat{H}^{(2)}& \hat{H}_{(1)} & \left(\hat{H}^{(0)} -\hbar \omega\right)&  \cdots \\ \Ddots & \vdots  &  \vdots  & \vdots   &  \ddots
\end{array} \right) \left(
\begin{array}{c} \vdots  \\  \ket{\phi^{(-1)}_{a}} \\ \ket{\phi^{(0)}_{a}} \\ \ket{\phi^{(1)}_{a}} \\ \vdots
\end{array} \right) = \varepsilon_a \left(
\begin{array}{c} \vdots  \\  \ket{\phi^{(-1)}_{a}} \\ \ket{\phi^{(0)}_{a}} \\ \ket{\phi^{(1)}_{a}} \\ \vdots
\end{array} \right).
\label{EQ:floqMatrix}
\end{equation}
\end{widetext}
In practice, solving the problem in nontrivial situations requires different approximations depending on the driving regime as, e.g., rotating wave-approximation or truncation of the Floquet-Hilbert space.

Here we obtain the Floquet solutions using a different calculation scheme. We start directly from Eq.\eqref{EQ:Htime} and consider the equation as an eigenvalue problem for the Floquet operator
\be
 H_{\rm F} \ket{\Psi_a(t)}= \tilde{E} \ket{\Psi_a(t)},
 \label{EQ:EigvFloquetNew}
\ee
where only the case of vanishing $\tilde{E}$ is physically relevant to solve the original problem.  Differently from Eq.\eqref{EQ:floquetQES}, here the eigenstates do not need to be time-periodic. Notice that if the physical time degree of freedom, $t$, is regarded as an extra spatial coordinate, $\tilde{x}$,  the Floquet operator can be then written as $H_{\rm F}= H(\tilde{x})+p_{\tilde{x}}$, with $p_{\tilde{x}}\smeq -\ci\hbar \partial_{\tilde{x}}$ the momentum along the new spatial dimension. The $H_{\rm F}$ can be interpreted as a Hamiltonian which is
periodic in the coordinate $\tilde{x}$. Such a Hamiltonian would determine the dynamics along an artificial time, $\tilde{t}$, according to the Schroedinger equation
\be
\left(H(\tilde{x})+p_{\tilde{x}}-\ci\hbar \frac{d}{d\tilde{t}} \right)  \ket{\Psi(\tilde{x},\tilde{t})}=0
\label{EQ:HF0}
\ee
where, as in Eq.\eqref{EQ:Htime}, the dependence of $\ket{\Psi}$ and $H$ on the spatial and spin degrees of freedom is not made explicit. Since $H_{\rm F}$ does not actually dependent on the artificial time $\tilde{t}$, one arrives to Eq.\eqref{EQ:EigvFloquetNew} by the standard procedure, writing $\ket{\Psi(\tilde{x},\tilde{t})}\smeq {\rm e}^{-\frac{\ci}{\hbar} \tilde{E} \tilde{t}}\ket{\Psi(\tilde{x})}$ with $\tilde{E}$ the artificial energy. Additionally, we can apply the standard Bloch theorem to deal with the periodicity along the coordinate $\tilde{x}$ (the physical time) and obtain the quasienergies from the quasimomenta along $\tilde{x}$ of the solutions with zero artificial energy. These $\tilde{E}\smeq 0$ solutions are those that do not depend on the artificial time, something expected for the physical solutions of the original system described by Eq.\eqref{EQ:Htime}.

Our calculation scheme resorts to a discretization along $\tilde{x}$ as a mean to define a lattice problem. We keep the time units for $\tilde{x}$ despite it is regarded as an additional spatial coordinate. Relevant features due to the discrete nature of the model are identified  by comparison with the continuous case. To this aim, it is sufficient to discuss the case of a $\tilde{x}$-independent Hamiltonian:
\be
H_{\rm F} = \sum_\alpha E_\alpha \ketLR{\alpha}\braLR{\alpha} -  \ci\hbar \partial_{\tilde{x}},
\label{EQ:HF1}
\ee
where ${\alpha}$ labels the eigenstates of the Hamiltonian restricted to real-space, spin, etc., at any given position $\tilde{x}$. In the lattice version of this problem we work with the discrete values $\tilde{x}_i=i \Delta \tilde{x}$, with $\Delta \tilde{x}$ the lattice spacing. The derivative in Eq.\eqref{EQ:HF1} generates a hopping term between first nearest neighbors sites along $\tilde{x}$ given by
\be
 - \ci \frac{\hbar}{2 \Delta \tilde{x}} \sum_\alpha  \ketLR{\alpha}\braLR{\alpha} \otimes \sum_i  \left( \ketLR{\tilde{{x}}_{i+1}}\braLR{\tilde{x}_{i}}-\ketLR{\tilde{x}_{i}}\braLR{\tilde{x}_{i+1}} \right)
\label{EQ:HF3},
\ee
all remaining terms in $H_{\rm F}$ are diagonal in $\tilde{x}_{i}$.

For the continuous model, after replacing $\ket{\Psi(\tilde{x},\tilde{t})}=\ket{\psi(\tilde{x})} {\rm e}^{-\frac{\ci}{\hbar} \tilde{E}\tilde{t}}$ in the Schroedinger equation of Eq.\eqref{EQ:HF0}, we arrive at the eigenvalue problem $H_{\rm F}\ket{\psi(\tilde{x})} =\tilde{E}\ket{\psi(\tilde{x})}$. The solutions for this $\tilde{x}$-independent system are classified by the quasimomenta $k_{\tilde{x}}$ and $\alpha$, with eigenfunctions $\ket{k_{\tilde{x}},\alpha} = {\rm e}^{\frac{\ci}{\hbar} k_{\tilde{x}} \tilde{x}} \ket{\alpha}$. Notice that we define the momentum $k_{\tilde{x}}$ with units of energy. The eigenenergies are
\be
\tilde{E}( k_{\tilde{x}},\alpha)= E_{\alpha} + k_{\tilde{x}}~~ \Rightarrow ~~ \left. k_{\tilde{x}}\right|_{\tilde{E}=0}=-E_{\alpha}.
\label{EQ:Econt}
\ee
After imposing the $\tilde{E}=0$, condition the dynamics in the artificial $\tilde{t}$ becomes irrelevant and we proceed by replacing $\tilde{x}$ with the physical time $t$. We obtain the well known solutions to the problem: $\ket{\Phi(t)} = {\rm e}^{-\frac{\ci}{\hbar} E_\alpha t} \ket{\alpha}$.

For the lattice model each $\alpha$ generates a one-dimensional tight-binding system which is solved by using the quasimomenta states $\ket{k_{\tilde{x}} }$, where $\ket{\tilde{x}_i }\smeq \frac{1}{\sqrt{2}} \sum_{\ve_{\tilde{x}}} {\rm e}^{\frac{\ci}{\hbar} k_{\tilde{x}} \tilde{x}_i} \ket{k_{\tilde{x}} }$.
We obtain the eigenenergies
\be
\tilde{E}_{\rm lat}( k_{\tilde{x}},\alpha)=  E_{\alpha} + \frac{\hbar}{\Delta\tilde{x}}\sin({\frac{\Delta\tilde{x}}{\hbar}k_{\tilde{x}}})~
\label{EQ:Elattice}
\ee
Notice that by imposing $\tilde{E}_{\rm lat}=0$ one finds two possible values of $k_{\tilde{x}}$: one of them is an artifact of the lattice construction and must be discarded. As the physical solutions (see Eq.\eqref{EQ:Econt}) have $\frac{\partial \tilde{E}_{\rm lat}}{\partial k_{\tilde{x}}}>0$ we must discard the $\tilde{E}_{\rm lat}\smeq 0$ solutions with \emph{negative velocity} along $\tilde{x}$. For the \emph{positive velocity} solution we find
\bea
k_{\tilde{x}}&=&-E_{\alpha}+\left(k_{\tilde{x}}-\frac{\hbar}{\Delta\tilde{x}}\sin(\frac{\Delta\tilde{x}}{\hbar}k_{\tilde{x}})\right) \\
&=&-E_{\alpha}+ k_{\tilde{x}}\left[\frac{1}{6}\left(\frac{\Delta\tilde{x}}{\hbar} k_{\tilde{x}} \right)^2 - \frac{1}{120}\left(\frac{\Delta\tilde{x}}{\hbar}k_{\tilde{x}}\right)^4 +\mathcal{O}(\frac{\Delta\tilde{x}}{\hbar}k_{\tilde{x}})^6\right] \notag.
\eea
The lattice approximation works well ($k_{\tilde{x}}\rightarrow -E_\alpha$) as long as $E_\alpha\ll\frac{\hbar}{\Delta\tilde{x}}$. This means that the lattice spacing, measured in time, must be sufficiently small to sample the wavelength associated to the relevant energy scale, $E_\alpha$. In other words, the bandwidth $2\frac{\hbar}{\Delta\tilde{x}}$ in Eq.\eqref{EQ:Elattice}  must be sufficiently large for a linear approximation of the sine function when $|k_{\tilde{x}}|< |E_\alpha|$. Besides, for a bandwidth smaller than $E_\alpha$ only evanescent $\tilde{E}_{\rm lat}=0$ solutions would appear. In that case, one would miss the (physically relevant) traveling solution that must exist as $H_{\rm F}$ does not have quadratic terms of momentum along $\tilde{x}$ (namely, time $t$).

In Floquet systems, $H(\tilde{x})$ is periodic along $\tilde{x}$. In this case, there are three relevant energy scales that must be much smaller than $\frac{\hbar}{\Delta\tilde{x}}$ in the lattice approximation. First, at each position $\tilde{x}_i$, the Hamiltonian can be diagonalized along the remaining (actual) dimensions with eigenenergies $E_\beta(\tilde{x}_i)$, where $\beta$ labels the instantaneous eigenstates.\footnote{Here $\beta$ is an index that may point to different states at each $\tilde{x}_i$.} The energy $E_\beta^{\max} \equiv \max_{\beta,\tilde{x}_i}|E_\beta(\tilde{x}_i)|$ must be smaller than $\frac{\hbar}{\Delta\tilde{x}}$. Second, given a driving term with frequency $\omega$ and assuming a description up to $N_{\rm max}$-photon processes, then $\frac{\hbar}{\Delta\tilde{x}}\gg N_{\rm max} \hbar\omega$. Finally, the driving amplitude, $\hbar \Omega$, in units of energy, must be kept smaller than the tight-binding bandwidth. In summary, the condition
\be
\frac{\hbar}{\Delta\tilde{x}}\gg \max\{N_{\rm max} \hbar\omega,~\hbar \Omega,~ E_\beta^{\max}\}
\label{EQ:condition}
\ee
guaranties an accurate tight-binding description of the Floquet problem.

The lattice construction, since it is time-independent along $\tilde{t}$, can be solved using the superlattice scheme discussed in Sec.\ref{SC:supl}. This allow us to find the Floquet quasienergy states directly from Eq.\eqref{EQ:hamil2}. For simplicity, from now on the physical time is referred to as $t$ or just time (keeping in mind that it can be interpreted as the additional spatial coordinate $\tilde{x}$). We proceed by discretizing a period of the driving potential in $P\ge 2$ time intervals during which the excitation is considered as constant. We assume that all intervals have the same duration $\Delta t$ and $T=P \Delta t$. [Time intervals of variable duration can be easily introduced without significant changes.] The instantaneous Hamiltonians at times $t_i=i \Delta t$ are $\{\hat{H}(t_1),\hat{H}(t_2),\cdots,\hat{H}(t_P)\}$, each of them corresponding to a mesoscopic system modeled by a $N$-dimensional lattice.\footnote{This includes the case of translational invariant systems subject to a translational invariant driving: For each value of the wavevector $\mathbf{k}$ one must solve an effective system described by a finite set of effective-lattice sites.} By discretizing the derivative operator in Eq.\eqref{EQ:hamil2} [see Eq.\eqref{EQ:HF3}], we derive a matrix representation for the operator $H_{\rm F}$:
\begin{equation}
 \left(
\begin{array}{ccccc}  \ddots & \vdots  &  \vdots  & \vdots   &  \Ddots \\ \cdots & \hat{H}(t_P)  & \ci\frac{\hbar}{2\Delta t} \eins_{N} & \zeins_{N} & \cdots
\\ \cdots   &  -\ci\frac{\hbar}{2\Delta t} \eins_{N} & \hat{H}(t_1) & \ci\frac{\hbar}{2\Delta t} \eins_{N} & \cdots
 \\ \cdots &  \zeins_{N} & -\ci\frac{\hbar}{2\Delta t} \eins_{N} & \hat{H}(t_2) &  \cdots \\ \Ddots & \vdots  &  \vdots  & \vdots   &  \ddots
\end{array} \right) \left(
\begin{array}{c} \vdots  \\  G(t_0) \\ G(t_1) \\ G(t_2) \\ \vdots
\end{array} \right) =  \left(
\begin{array}{c} \vdots  \\   0 \\ 0 \\ 0 \\ \vdots
\end{array} \right).
\label{EQ:HfullTime}
\end{equation}
Here, $G(t_i)=(g_{1}(t_i),g_{2}(t_i), \cdots,g_{N}(t_i))^T$ is a $N$-dimensional vector representing the amplitudes at time $t_i$ of the solution $\ket{\Psi}$ on the lattice's sites. The instantaneous Hamiltonians $\hat{H}(t_i)$ are written in the same basis.

We then find that periodically driven systems can be approached and solved as a regular spatial superlattices (see Sec.\ref{SC:supl}) by substituting
\bese
\bea
\Delta t &\leftrightarrow&  a_0 \\
G(t_i) &\leftrightarrow& G_i ~~,~\forall i \\
\ci\frac{\hbar}{2\Delta t} \eins_{N} &\leftrightarrow& \hat{T}_{i,i+1}~~,~\forall i \\
\hat{H}(t_i) &\leftrightarrow&  \hat{H}_i~~,~\forall i \\
\widetilde{\lambda}\equiv {\rm e }^{\ci  k_x P a_0} &\leftrightarrow& \widetilde{\lambda}\equiv {\rm e }^{-\ci  \ve P \Delta t / \hbar}.
\eea
\eese
As the right hand side of Eq.\eqref{EQ:HfullTime} is zero, \emph{only} the $\tilde{E}=0$ characteristic matrix [either $\mathcal{M}(0)$ or $\widetilde{\mathcal{M}}(0)$] is relevant to solve the time-dependent problem. The Floquet quasienergies are obtained as
\be
\ve= \frac{\hbar}{P \Delta t} \mathrm{Arg} \widetilde{\lambda}_l,
\label{EQ:Quasienergies}
\ee
from the eigenvalues $\widetilde{\lambda}_l$ of traveling solutions with positive velocity; i.e., $\widetilde{\lambda}_l\smeq 0$ and $v_l> 0$, see Eq.\eqref{EQ:velocitySuperlattice}.

Finally, the hopping amplitude is $\ci\hbar P/2 T$ (after $T\smeq P \Delta t$), increasing with the amount of sites per period. While a larger $P$ (smaller $\Delta t$) would produce a better approximation according to Eq.\eqref{EQ:condition}, the computational cost for calculating $\mathcal{M}(0)$ can be large.
Such a trade-off between precision and computational cost is also present in the frequency space treatment of Eq.\eqref{EQ:HfullTime} as the more Fourier modes one includes before truncation the better the final solution.

\section{Case study: Topological superconductors}
\label{SC:Examples}
\subsection{Single- and multi-layer superlattices: \\Topological superconductivity in multimode Rashba wires}
\label{SC:EgQWire}

We start by illustrating the standard eigenvalue method for static systems. As an example, we study the case of a quantum wire subject to SOC in the vicinity of a superconductor. In this situation $s$-wave superconducting pairing is induced in the wire by proximity effect. By introducing an additional Zeeman field, the system
can develop a topological superconducting (TS) phase,\cite{SauDasSarma2010,*Alicea2010prb,*OregRefaelvonOppen2010} behaving as an effective spinless p-wave superconductor. At the edges of a TS wire, zero-energy Majorana bound states appear as mid-gap solutions of the Bogoliubov-deGennes (BdG) equation. These solutions are--- by virtue of electron-hole symmetry--- their own ``antiparticles". A pair of Majorana fermions (MFs) apart from each other can be used to encode a qubit protected from local perturbations, highly interesting for quantum information and quantum computation purposes.\cite{Ivanov2001prl,Kitaev2001,Stern2004prb,TQCrmpNayak,AliceaNatPhys2011}

Single-mode (or quasi one-dimensional) quantum wires have been extensively studied, including interaction effects\cite{Stoudenmire2011} and quasiperiodic lattice modulations\cite{Tezuka2012,DeGottardi2013PRL,AdagideliWimmer2013}.
Here we consider a multimode quantum wire. We apply the eigenvalue method to visualize the dispersion relations for particular parameter settings to obtain the effective superconducting gap in order to produce, whenever possible, MFs from the zero-energy evanescent solutions. This supports and extends previous results for multiband wires.\cite{PotterLee2010,*PotterLee2011,Lutchyn2011,Tewari2012,Serra2013} We further apply the eigenvalue method to study the effect of a superlattice potential in the multimode Majorana wire.

\subsubsection{Normal quantum wire}

We first discuss a multichannel quantum wire in absence of the superconducting pairing. The setup is depicted in Fig.\ref{FG:fig2}(a). We obtain the corresponding band structure by assuming translation invariance along the $x$ direction and confinement in the $y$ direction [hard-wall quantum well of width $W$ defined by a potential $V_{\rm W}(y)$]. An in-plane magnetic field is applied along the wire with a corresponding Zeeman energy $E_Z \equiv g\mu_B B_x/2$. The Hamiltonian, in continuous variables, reads
\be
H_{\rm N}=\frac{1}{2m^*}\hat{p}_x^2 -\frac{\alpha_\parallel}{\hbar}\hat{p}_x\sigma_y+\frac{\alpha_\perp}{\hbar}\hat{p}_y\sigma_x+ E_{\rm Z} \sigma_x+V_{\rm W}(y),
\label{EQ:Hcont}
\ee
where the SOC strengths $\alpha_\parallel$ and $\alpha_\perp$ selectively affect--- for studying the effect SOC in channel mixing--- the components involving the longitudinal and transversal linear momentum, respectively.

The eigensatates are plane waves and the transverse modes (labeled $n\smeq 1,2,...$) have energy offsets ($k_x\smeq0$):\cite{ReynosoUB07,*ReynosoUB08prb}
\be
E_{n,\sigma}(k_x=0,E_{\rm Z})=\frac{\hbar^2 \left(n \pi \right)^2}{2 m^*} + \frac{{\alpha_\perp}^2 E_\alpha}{{\alpha_\parallel}^2} +E_{\rm Z}\sigma
\label{EQ:En}
\ee
with $\sigma\smeq\{+,-\}$ (or $\{\uparrow_x,\downarrow_x\}$), the spin projection along the magnetic field axis and $E_\alpha\smeq {m^*{\alpha_\parallel}^2/\hbar^2 2}$. These levels are shown as a function of $E_{\rm Z}$ in Fig.\ref{FG:fig4}(a), labeled as $(n,\sigma)$. They become independent of the spin $\sigma$ for zero $E_Z$, with zero-field energy $E_n\equiv E_{n,\sigma}(k_x\smeq 0,E_{\rm Z}\smeq 0)$. [For vanishing $B_x$ and $\alpha_\perp$, and  away from $k_x\smeq 0$, the spin eigenstates point along the $y$-direction by virtue of the term $(\alpha_\parallel/\hbar)\hat{p}_x\sigma_y$.\cite{EtoHK05,ReynosoAndreev2012}] Notice that, as $E_2\smmi E_1$ does not depend on $\alpha_\perp$, the level crossings shown in Fig.\ref{FG:fig4}(a) for $E_Z\smeq \Delta E_{21}\equiv E_2\smmi E_1$ (relevant for the topological phase transitions discussed in next section) are present independently of the SOC strength.

To apply the method, we discretize Eq.\eqref{EQ:Hcont} arriving to a tight-binding model detailed in the Sec.\ref{AP:TBnormal}. We implement on-site and hopping Hamiltonians $\hat{H}_0$ and $\hat{T}_0$ based on  a discretization of $20$ sites across the wire (width $W=21 a_0$). Complex band structures are readily obtained by diagonalizing the matrix $\mathcal{M}(E)$ given in Eq.\eqref{EQ:TransMat} and sweeping over $E$. For illustration, in Fig.\ref{FG:fig2}(a) we show the results for first three transverse modes. Solutions with $\mathrm{Im}[k_x]\neq 0$ are evanescent and thus they are not contained in the $\mathrm{Re}[k_x]-E$ plane.

The dispersions $E(k_x)$ of traveling eigenstates follow from solutions with $\mathrm{Re}[k_x]=k_x$. The solutions at different $E$ can be further classified according to the transverse modes, velocity, spin properties, etc. In Fig.\ref{FG:fig2}(b) we present dispersion relations corresponding---as the complex band structure in Fig.\ref{FG:fig2}(a)---to the discretized version of Hamiltonian \eqref{EQ:Hcont}. For $\alpha_\perp\smeq 0$, the dispersion presents true crossings as both the magnetic field ($B_x$) and the parallel SOC cannot mix states of different transverse modes. Avoided crossings are induced, instead, when $\alpha_\perp\smeq \alpha_\parallel$. This is expected as the term $-\ci\alpha_\perp\partial_y\sigma_x$ mixes transverse modes of different parity (axially symmetric and antisymmetric) provided the spin components are mixed by $\sigma_x$.
In Fig.\ref{FG:fig2}(c) we show the velocity $v$ of each solution [classified as in Fig.\ref{FG:fig2}(b)] as a function of $E$. In the presence of a finite $B_x$, for solutions that lie in the vicinity of $k_x\smeq 0$, the function $E(v)$ is multi-valued for some bands. The $E(v)$ functions show clear evidence of the mixing induced by $\alpha_\perp$.

\subsubsection{Superconducting wire}

\begin{figure}[!t]
\includegraphics[width=0.44\textwidth]{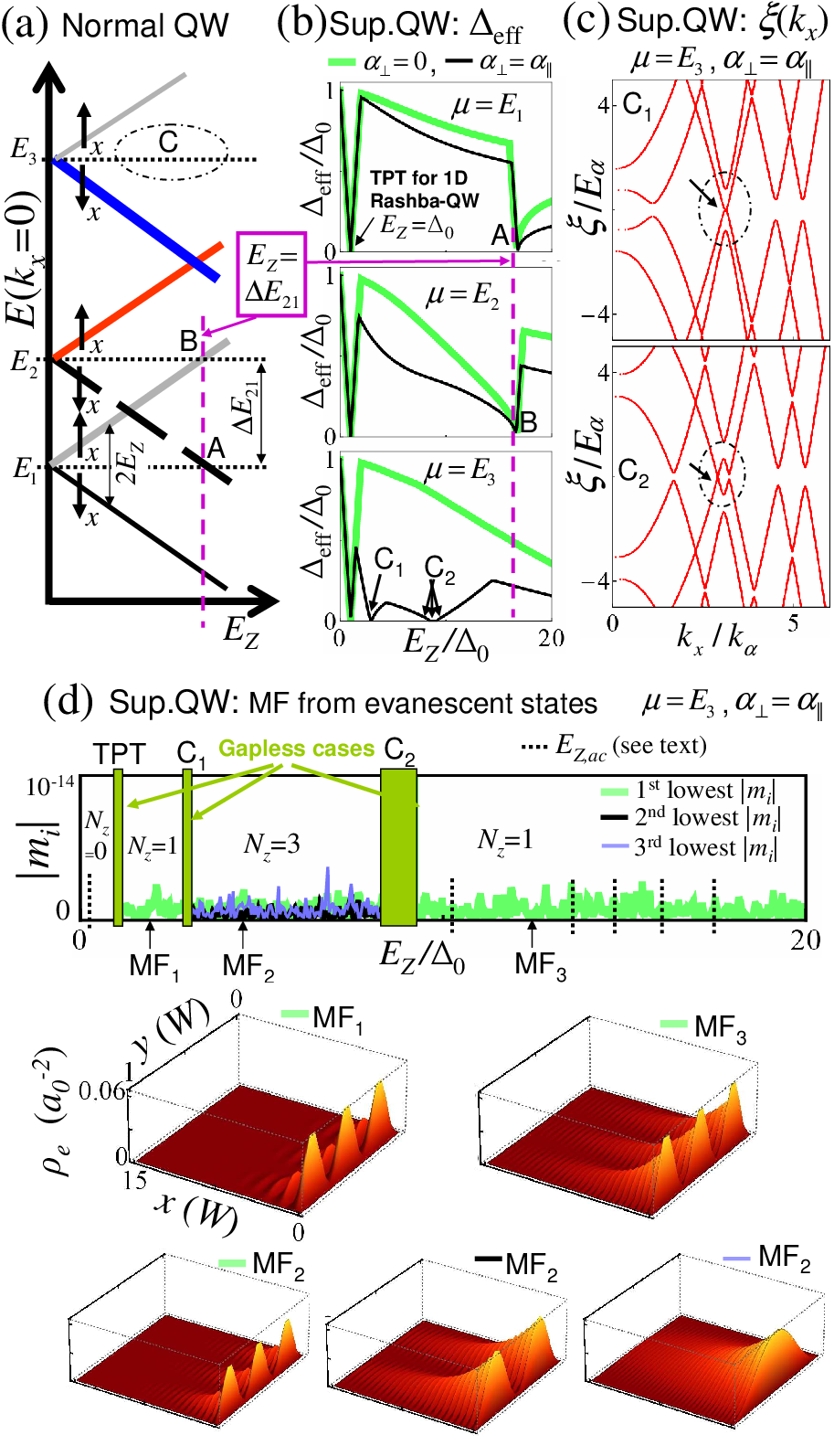}
\vspace{-0.5cm}
\caption{\label{FG:fig4} Superconducting multimode Rashba quantum wire: Effective energy gap, $\Delta_\mathrm{eff}$, dispersion relations, $\xi(k_x)$, and Majorana fermions constructed from evanescent modes. Parameters as given in Fig.\ref{FG:fig2}, i.e., $\alpha_\parallel$ is nonzero and fixed. (a) Normal case ($\Delta_0\smeq 0$): Sketch of the first three transverse modes $k_x=0$ energy levels as a function of the Zeeman energy $E_Z$: At zero field both $(n,\sigma)$ levels, $\sigma\smeq\uparrow_x,\downarrow_x$, for transverse mode mode $n$ fall at $E_n$. When the Zeeman energy is $\Delta E_{21}\equiv E_2\smmi E_1$ the $k_x=0$ energy of states $(1,\uparrow_x)$ [$(2,\downarrow_x)$] crosses $E_2$ [$E_1$]. (b) For  $\Delta_0\smeq 0.4 E_\alpha$ we show $\Delta_\mathrm{eff}$ for $\mu\smeq E_1$, $\mu\smeq E_2$ and $\mu\smeq E_3$, the realistic $\alpha_\perp\smeq\alpha_\parallel$ situation is compared with the $\alpha_\perp\smeq0$ case. The gap closes at $E_Z/\Delta_0\smeq1$ always: as expected from the topological superconductor TPT, and for $\mu\smeq E_{1,2}$ another gap closing is observed at $E_Z\smeq\Delta E_{21}$ in agreement with panel (a). In general $\alpha_\perp\smeq\alpha_\parallel$ reduces the effective gap (black line), for $\mu\smeq E_3$ $\Delta_\mathrm{eff}$ goes to zero at unexpected values of $E_Z$. Panel (c) shows $\xi(k_x)$ at the gap closing value $C_1$ and within the range $C_2$ being due to $\alpha_\perp$-induced mixing between the second and the first transverse modes. (d) For $\mu\smeq E_3$ we use the lowest eigenvalues of the evanescent-matrix to count the number of MF bound states, $N_z$. The boxes indicate gapless conditions in which the evanescent matrix is nonphysical. As expected $E_Z\smeq\Delta_0$ separates the low field trivial superconductor region ($N_z\smeq 0$) from the topological superconductor region ($N_z\smeq 1$). In between $C_1$ and $C_2$ we find $N_z\smeq 3$, this is odd and thus it leads to a protected topological superconductor. For representative values of $E_Z$ we show the local electronic density, $\rho_e(x,y)$, of the obtained MF bound states.}
\vspace{-0.1cm}
\end{figure}

We now proceed by including the $s$-wave superconducting pairing through a standard Bogoliubov-deGennes (BdG) equation approach.\cite{deGennesBook} This introduces two relevant parameters: the chemical potential $\mu$ and the pairing's amplitude $\Delta_0$. The BdG equation reduces to $(\hat{H}_{BdG}-\xi)\ketLR{\Psi} =0$, with $\xi$ the energy measured from the chemical potential ($\xi\equiv E-\mu$, reserving the usual notation $\ve$ for Floquet quasienergies).
The corresponding lattice model is presented in Sec.\ref{AP:TBsup}.\cite{Cuevas96} After a generalization of the $\hat{H}_0$ and $\hat{T}_0$ to the particles and holes as treated in the BdG equation, we apply the eigenvalue method to obtain the corresponding $\mathcal{M}(\xi)$. Results are presented in Fig.\ref{FG:fig4}.

We sweep $\xi$ to obtain the effective gap, $\Delta_\mathrm{eff}$, as a function of the Zeeman energy in the range $E_{\rm Z}<20\Delta_0$. $\Delta_\mathrm{eff}$ is defined as the minimum $\xi$ for which $\mathcal{M}(\xi)$ has at least one traveling solution with eigenvalue $|\lambda_l|=1$. For $|\xi|<\Delta_\mathrm{eff}$ only evanescent solutions exist. In Fig.\ref{FG:fig4}(b) we set the chemical potential $\mu\smeq \{E_1,E_2,E_3\}$, corresponding to the spin-degeneracy energies of the first three transverse modes at $k_x=0$ for the normal system [i.e., vanishing $E_{\rm Z}$ and $\Delta_0$, see Fig.\ref{FG:fig4}(a)]. These choices of $\mu$ are the most favorable to observe the topological superconducting phase as they lead to larger $\Delta_\mathrm{eff}$.
First, we discuss features independent of $\alpha_\perp$. We observe the topological phase transition (TPT)--- from trivial to topological superconductivity--- at $E_{\rm Z}\smeq \Delta_0$. At the topological transition, $\Delta_\mathrm{eff}$ grows linearly with $E_{\rm Z}$ as it is dominated by the small $k_x$ physics.\cite{SauDasSarma2010,*Alicea2010prb,*OregRefaelvonOppen2010} For gaps appearing at finite $k_x$, $\Delta_\mathrm{eff}$ typically decreases with $E_{\rm Z}$ up to a point at which linear dependence on $E_{\rm Z}$ is recovered, producing additional closings of the gap. This happens, for instance, in the cases $\mu\smeq \{E_1,E_2\}$ as we observe the closing of the gap at $E_{\rm Z}\smeq \Delta E_{21}$, which is related to the level crossing indicated in Fig.\ref{FG:fig4}(a). For $E_{\rm Z}$ larger than the gap closing at $\Delta E_{21}$ the superconductor is topologically trivial: Two Majorana fermions would exist at each edge of the system (in a pure and perfect finite sample) but robustness to local perturbations is lost, due to mixing among the even number of solutions localized at the same edge, and no Majorana fermions survive in real samples.\cite{PotterLee2010,*PotterLee2011,Lutchyn2011,Tewari2012,Serra2013}

The introduction of a finite $\alpha_\perp$ produces a reduction of the effective gap in regions where $\Delta_\mathrm{eff}$ does not follow a linear dependence on $E_{\rm Z}$. This means that $B_x$ degrades superconductivity more efficiently in the presence of the transverse mode mixing induced by $\hat{p}_y \sigma_x$, particularly at large $k_x$. More remarkable, for $\mu\smeq E_3$ we observe gap closings at values of $E_{\rm Z}$ that are unrelated to level crossings at $k_x\smap 0$ (see region $C$ in Fig.\ref{FG:fig4}(a)): As shown in Fig.\ref{FG:fig4}(b), the Zeeman energy $E_{\rm Z}$ at point $C_1$ and in the range $C_2$ produces $\Delta_\mathrm{eff}\smeq0$. Figure \ref{FG:fig4}(c) shows the dispersion relations, $\xi(k_x)$, obtained with the eigenvalue method for $E_{\rm Z}$ at $C_1$ and within $C_2$. The $\xi(k_x)$ are depicted for $k_x\geq 0$ as they are even functions of $k_x$. We observe that the gap closings are related to first and second mode mixing at large values of $|k_x|$. Indeed, the region between $C_1$ and $C_2$ is in a topological superconducting phase leading to protected Majorana edge states (as we prove below by calculating a topological invariant).

\subsubsection{Topological phase identification: counting Majorana end states}

In a superconducting phase, i.e., away from gapless conditions, only evanescent states exist for $\xi<\Delta_\mathrm{eff}$. We can obtain these states by applying the complex band structure method: This states are indeed useful to identify the topology of the superconducting phase by counting the number of MFs that would appear in a sample edge. We then impose a wire termination at $x=0$ (with vacuum for $x<0$), and proceed to search for independent solutions satisfying the boundary condition using evanescent states decaying for $x>0$. As MFs are bound states at $\xi\smeq0$, we only need to work with the unbounded solutions extracted from $\mathcal{M}(0)$. We label these evanescent states ($\mathrm{Im}[k^{(l)}_x]<0$) as $\ketLR{\Phi_l}$ with $l\smeq1,2,..,N$. Note that $\ketLR{\Phi_l}$ is a $N$-dimensional vector having amplitudes along the $j\smeq1,2,..,N$ effective sites in a given layer. Any $\xi\smeq0$ solution of the open wire problem must have the form $\ketLR{\Phi(x)} \equiv \sum_l  f_l \ketLR{\Phi_l}{\rm e}^{\ci k^{(l)}_x x }$ with the coefficients $f_l$ chosen to satisfy the boundary condition $\ketLR{\Phi(x\smeq0)}\smeq\sum_l  f_l \ketLR{\Phi_l}\smeq \ketLR{0_N}$, with $\ketLR{0_N}$ the $N$-dimensional zero ket. By projecting the latter equation on state $\braLR{\Phi_h}$ with  $l\smeq1,2,..,N$, we find $N$ algebraic equations $\sum_{l}  f_l \braket{\Phi_h|\Phi_l}\smeq 0$. For convenience, we introduce the \emph{evanescent matrix} $\bar{F}$ with elements $\bar{F}_{h,l}\smeq \braket{\Phi_h|\Phi_l}$  and the vector $\vec{f}\equiv (f_1,f_2,..,f_N)^T$. The boundary condition reads\cite{Serra2013}
\be
\bar{F} \cdot \vec{f}= \vec{0}.
\ee

The problem reduces to find the kernel of the evanescent matrix, we call $N_z$ to the dimension of the null subspace. Each of those $N_z$ zero eigenvalues produces a MF solution ${\Phi_{\mathrm{MF}}(x,y,\tau,\sigma)}\smeq \sum_l  f^{\mathrm{MF}}_l \braket{y,\tau,\sigma|\Phi_l}{\rm e}^{\ci k^{(l)}_x  x }$. The MF solution has an exponential dependence on $x$ while the $N$ components of $\ketLR{\Phi_l}$ encode the dependence on $y$ and spin $\sigma$ in the electron-hole blocks $\tau\smeq\{e,h\}$. We define the electronic probability density of a MF solution as $\rho_e(x,y)\equiv \sum_\sigma \left|{\Phi_{\mathrm{MF}}(x,y,\tau=e,\sigma)}\right|^2$.

By using the method, we are able find a set of eigenvalues that can be considered zero within numerical noise. These eigenvalues are well separated (by around $10$ orders of magnitude) from the next-closest to zero eigenvalues which are not associated to solutions satisfying the boundary condition. The method is therefore well suited to calculate the topological number associated to a given superconducting phase. In Fig.\ref{FG:fig4}(d) we plot the absolute values of the first three closest to zero eigenvalues of $\bar{F}$, $m_i$,  as a function of $E_{\rm Z}$ for the quantum wire discussed above, we choose $\mu\smeq E_3$ and $\alpha_\perp\smeq\alpha_\parallel$. We obtain $N_z$ by counting the number of $m_i$ that are zero (within numerical precision). For $E_Z<\Delta_0$ we find a trivial topological phase since, as expected, $N_z\smeq 0$ and no MFs can exist at the wire's end. For larger $E_Z$, we find an odd-parity $N_z$ indicating the development of a topological superconducting phase, $N_z=1$: In the presence of disorder, for odd-parity $N_z$, only one MF would survive at the wire's end (remaining at zero energy) since the electron-hole symmetry of the BdG equation forces the states to appear as $-E,E$ pairs. For the same reason even-parity $N_z$ phases on the other are not protected against the mixing induced by local perturbations. Notice that in the region between the gap closings $C_1$ and $C_2$ (related to mode mixing due to $\alpha_\perp$) we find $N_z=3$, i.e., this is a topological phase. Occasionally, for some isolated values of $E_{\rm Z}$ [see dotted lines, $E_{\rm Z,ac}$ in Fig.\ref{FG:fig4}(d)] we find that an additional eigenvalue of $\bar{F}$ tend to vanish. We do not discuss these situations further as they are not physically significant: (i) $N_z$ is unchanged across a given $E_{\rm Z,ac}$ since there is no qualitative change in $\Delta_\mathrm{eff}$ and (ii) the additional MF only survives in a region of zero measure (within the $E_Z$-parameter space), and thus they are out of experimental reach.

In Fig.\ref{FG:fig4}(d) we also plot the electron density $\rho_e(x,y)$ for the typical MF in each region. Interestingly, in the region of $N_z\smeq 1$ we find the signature of the third transverse mode in $\rho_e(x,y)$ (according to the fact that $\mu\smeq E_3$). In the region $N_z\smeq3$, instead, the three MF solutions are associated to the all $3$ transverse modes. The latter result is due to the transverse mode mixing between $C_1$ and $C_2$: For $\alpha_\perp\smeq0$, this region turns to be a $N_z\smeq 1$ phase and $\rho_e(x,y)$ shows the characteristic of the third
transverse mode only (not shown).

\begin{figure}
\includegraphics[width=0.45\textwidth]{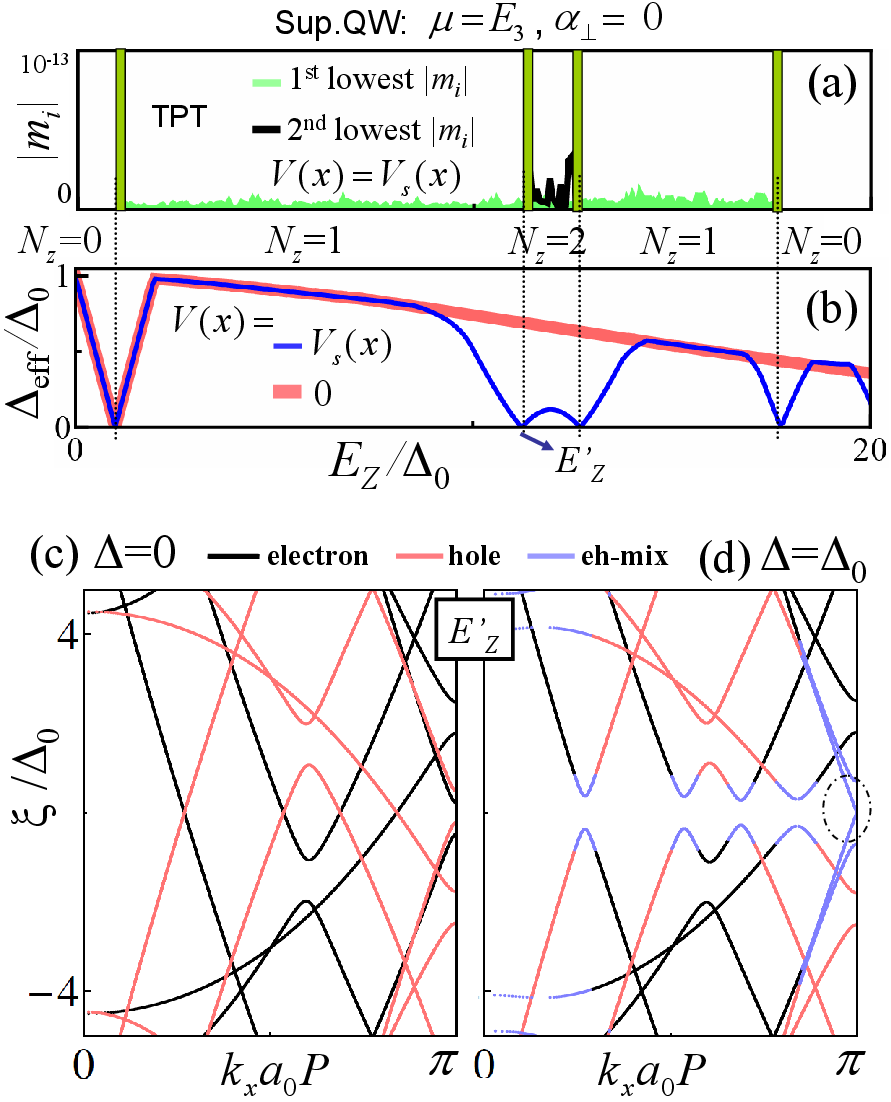}
\vspace{-0.4cm}
\caption{\label{FG:fig5} Superconducting multimode Rashba quantum wire in the presence of a superlattice potential: Effective energy gap, $\Delta_\mathrm{eff}$, dispersion relations, $\xi(k_x)$, and Majorana fermions constructed from evanescent modes. Parameters are as Fig.\ref{FG:fig2} and Fig.\ref{FG:fig4} with $\alpha_\perp\smeq0$ and $\mu\smeq E_3$. The superlattice potential is $V_s\smeq 3\Delta_0 \cos (2\pi x/(P a_0))$ with $P a_0\smeq W/2$. (a) We extract $N_z$, the number of MF bound states at an open boundary in the clean sample, from the number of eigenvalues of the evanescent-matrix which are numerically zero. The superlattice potential induces non-protected regions as $N_z$ becomes even (2 and 0) in regions that otherwise would be topological ($N_z\smeq 1$ for $V(x)\smeq 0$). (b) Due to the superlattice potential $V(x)\smeq V_s(x)$ the effective gap $\Delta_\mathrm{eff}$ goes to zero precisely at the $E_Z$ values in which $N_z$ [in (a)] change from even to odd. (c) We show the dispersion relations $\xi(k_x)$ for the value $E_Z'$ of the band closing shown in panel (b). [As sketched in Fig.\ref{FG:fig1}(b), the dispersions are reconstructed by sweeping the energy, therefore, the separation between calculated points is greater the smaller is $|\partial_{k_x}\xi(k_x)|$; such separation between plotted points (see for example branches at $\xi/\Delta_0 \approx \pm4$) does not mean that the true dispersion is discontinuos.]  We choose $\Delta\smeq 0$ (left panel) and demonstrate the ability of the method to identify the electron and the hole branches. For $\Delta\smeq \Delta_0$ (right panel) we identify the regions where the superconducting pairing mixes the electrons and hole solutions by using the concept of electron-hole polarization, $P_{eh}$ (see text). The band closing appears as a Dirac cone at $k_x a_0 P \smeq \pi$ , precisely at the boundary of the first Brillouin zone: Notice that the other half of the Dirac cone follows from the $k_x<0$ solutions, which are not shown as here they are a mirror image of the $k_x>0$ solutions.}
\vspace{-0.1cm}
\end{figure}

\subsubsection{Superconducting superlattice}

We add a spatial dependent potential $V(x)$ to the multichannel quantum wire discussed above. We choose the $P a_0$-periodic potential ($V(x)\smeq V(x \smpl P a_0)$)
\be
V(x)= V_s(x) =  V_0 \cos[2\pi x/(P a_0)]
\ee
with amplitude $V_0\smeq 3\Delta_0$. The associated lattice Hamiltonian in Sec.\ref{AP:TBsup} is used compute the characteristic matrix $\mathcal{M}(\xi)$ in order to solve the eigenvalue problem in Eq.\eqref{EQ:SupLatM}. Since $P>1$, the effective Hamiltonian needs to be recalculated at each value of $\xi$ by following Eqs.\eqref{EQ:SupLatAlg0}-\eqref{EQ:SupLatAlg2}, after replacing $E$ with $\xi$. In Figs.\ref{FG:fig5}(a) and \ref{FG:fig5}(b) we show $N_z$ and $\Delta_\mathrm{eff}$ as a function of $E_{\rm Z}$, respectively, for $\mu\smeq E_3$ and $\alpha_\perp\smeq0$.
If we had set $V(x)\smeq 0$, in the explored region of $E_{\rm Z}$, $N_z$ would change only at the $E_Z\smeq \Delta_0$ located TPT, from $N_z\smeq 0$ to $N_z\smeq 1$ (not shown). On the other hand, the superlattice potential $V_s(x)$ induces three extra $\Delta_\mathrm{eff}\smeq 0$ conditions at $E_Z>\Delta_0$. This induces regions of $E_Z$ in which the superconductor becomes trivial as $N_z$ is no longer $1$ but instead even: $N_z=2$ or $N_z=0$. The $N_z=2$ region can be understood as due to the addition (in the pure system) of one Majorana fermion associated with a unique TPT produced at finite $k_x$. At $E_Z \smeq E_Z'$, i.e., for the critical condition of the $N_z$ $1$ to $2$ TPT, in Fig.\ref{FG:fig5}(d) we show that $\xi(k_x)$ forms a Dirac cone-like dispersion relation closing exactly at the boundary of the first Brillouin zone: Note that the solutions of $k_x<0$ (not shown) form the other half of this same Dirac cone. This justifies that only a single Majorana is added as opposite to the $N_z$ $1$ to $3$ TPT at $C_1$ in Fig.\ref{FG:fig4}(c) where two extra Dirac cones close due to SOC mixing as the one at $k_x>0$ is not equivalent to the Dirac cone at $k_x<0$.

To illustrate the abilities of the method, we depict the dispersion $\xi(\ve)$ according to the relative electron and hole component. For each state, we compute the electron-hole polarization $P_{eh}\smeq (A_e-A_h)/(A_e+A_h)$ with $A_e$ and $A_h$ the electron and hole weight, respectively, as defined in Sec.\ref{AP:TBsup}. In Fig.\ref{FG:fig5}(c) we present results for $\mu\smeq E_3$ and $\alpha_\perp\smeq0$ with \emph{vanishing} superconducting pairing, verifying that the solutions correspond to decoupled electron and holes states related by particle-hole symmetry around $\xi\smeq0$.
The superconducting pairing is turned on in Fig.\ref{FG:fig5}(d), where we plot the solutions corresponding to $P_{eh}>0.9$ (electron-like), $P_{eh}<-0.9$ (hole-like), and $|P_{eh}|<0.9$ (electron-hole-like). This allows us to discriminate the regions affected by the $s$-wave superconducting pairing term. Importantly, this includes the region of the TPT related to the Dirac cone at the boundary of the first Brillouin zone.

\begin{figure*}
\includegraphics[width=0.95\textwidth]{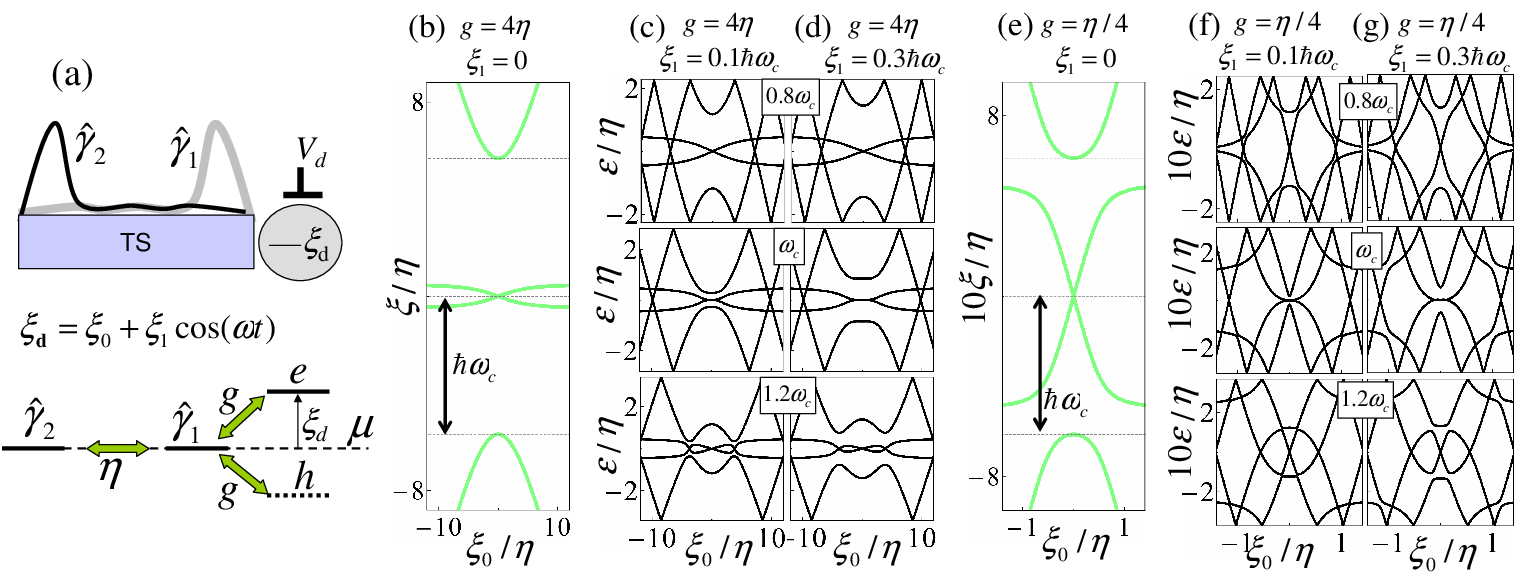}
\vspace{-0.1cm}
\caption{\label{FG:fig6} Microwave excited quantum dot in contact with a topological superconductor. (a) Sketch of the device and relevant states for the low energy ($|\xi| \ll\Delta_\mathrm{eff}$) description: the two Majorana fermions $\hat{\gamma}_1$ and $\hat{\gamma}_2$ and a single available state in the quantum dot. The energy of the dot, $E_d$, is controlled by the gate voltage $V_d$, measured from the chemical potential it is $\xi_d\equiv E_d\smmi \mu \smeq \xi_0 \smpl \xi_1 \cos{\omega t}$. Mixing between the MFs arises (matrix element of amplitude $\eta$) due to finite size of the TS region; also there is overlap between $\hat{\gamma}_1$ and the electron or hole in the quantum dot (matrix element of amplitude $g$).
Panels (b) and (e) show the energy levels in absence of the driving ($\xi_1\smeq0$) for $g\smeq4\eta$ and $g\smeq\eta/4$, respectively. At $\xi_0\smeq 0$ the Majorana condition is satisfied and poor man's Majorana fermions are generated (see text), the tuning condition is linearly sensitive to the dot energy. The energy of the high energy states at $\xi_0\smeq 0$ defines the critical frequency $\omega_c$.
For $g\smeq4\eta$ [$g\smeq4\eta/c$] we show the Floquet quasienergies of the driven system for driving amplitude $\xi_1\smeq\hbar \omega_c/10$ in (c) [(f)] and for $\xi_1\smeq 3\hbar \omega_c/10$ in (d) [(g)]. The Floquet Majorana condition at $\xi_0\smeq 0$ is also satisfied but the sensitivities to $\xi_0$ of the tunning condition is reduced when the dot is driven at the critical frequency: increasing the microwave amplitude $\xi_1$ at $\omega\smeq\omega_c$ reduces such sensitivity.}
\vspace{-0.1cm}
\end{figure*}

\subsection{Floquet Physics: Microwave excited quantum dot in contact with a topological superconductor}
Confined electrons in quantum dots (QDs) is one of the most interesting candidates for devising qubits in condensed matter systems.\cite{Petta2005,HansonReview} Recently, it has been noticed that interesting physics can emerge from the embedding of semiconducting QDs into TS systems. Such a combination can provide new ways to access TS properties via transport and the opportunity to devise hybrid spin-qubit/Majorana fermions quantum-computation and information-storage schemes.\cite{FlensbergPRL2011,Harold2011PRB,Leijnse2011,*Leijnse2011PRL} Even for QDs systems in contact with trivial superconductors, Majorana fermions (though not fully protected) can arise in static,\cite{SauPoor2012,Leijnse2012poor,Fulga2013,Brunetti2013} and Floquet situations.\cite{LiPoorMan2013} Here, we apply our method to study the case of a single QD placed the end of a TS wire such that the dot's levels $\xi_\mathrm{d}$ (measured from the superconductor's chemical potential $\mu$) are subject to a microwave driving as sketched in Fig.\ref{FG:fig6}(a).

For modeling the TS finite wire we neglect the quasiparticle states with $\xi\geq\Delta_\mathrm{eff}$: We work with the two MFs expected at each end of the wire, $\hat{\gamma}_1$ and $\hat{\gamma}_2$. As their energy lies exactly at the chemical potential ($\xi\smeq0$), they are protected by the gap $\Delta_\mathrm{eff}$. [We adopt the convention $\{\hat{\gamma}_i,\hat{\gamma}_j\}\smeq \delta_{ij}$ and thus $\hat{\gamma}_i^2=1/2$]. However, due to finite length of the wire, a mixing term  $-\ci \eta/2 \hat{\gamma}_1 \hat{\gamma}_2$  ($\eta\in\mathbb{R}$) arises between the two MFs. As a consequence, the actual solutions are $\hat{f}=2^{-\frac{1}{2}}(\hat{\gamma}_1 +\ci \hat{\gamma}_2)$ (with energy $+\eta/2$) and  $\hat{f}^\dagger=2^{-\frac{1}{2}}(\hat{\gamma}_1 -\ci \hat{\gamma}_2)$ (with energy $-\eta/2$), where the operator $\hat{f}$ fulfills usual fermionic anticommutation relations.

Regarding the QD, we assume that only one single electronic level is relevant at low energies ($|\xi_{\rm d}=E_{\rm d}-\mu| \ll \Delta_\mathrm{eff}$, with $E_{\rm d}$ the dot's energy) due to strong Coulomb repulsion with a second electron. Furthermore, the presence of the external magnetic field polarizes the electron spin state in the dot along the direction of the external field. The dot's occupancy changes from 0 to 1 when $\xi_\mathrm{d}$ changes sign from positive to negative. We now introduce a periodic driving of the dot's energy $\xi_\mathrm{d}(t)\equiv \xi_0\smpl\xi_1 \cos(\omega t)$.
As shown in Fig.\ref{FG:fig6}(a), the QD is contacted with the wire's end associated to the $\hat{\gamma}_1$ MF. This introduces a hopping term $g\hat{\gamma}_1\hat{d}+ h.c.$ in the Hamiltonian. By assuming that $\Delta_\mathrm{eff}$ in the TS is much larger than $g$, $\eta$, $\xi_0$, $\xi_1$ and $\hbar\omega$, the Hamiltonian reduces to
\be
\hat{H}(t)= \xi_\mathrm{d}(t) \hat{d}^\dagger \hat{d}-\ci\eta/2 \hat{\gamma}_1\hat{\gamma}_2 - g \hat{\gamma}_1 \hat{d} -g^* \hat{d}^\dagger \hat{\gamma}_1,
\ee
where $\hat{d}$ is the electronic annihilation operator in the dot. To treat this problem, one may rewrite the MF operators in terms of the fermion $\hat{f}$ and use the many-particle states $\ketLR{n_d,n_f}$ as a basis (with $n_f$ and $n_d$ the eigenvalues of the number operators, $\hat{f}^\dagger\hat{f}$ and $\hat{d}^\dagger\hat{d}$, respectively). Here, instead, we obtain the single-particle excitations of the system. This simplifies the finding of Majorana solutions: For the static case ($\xi_1\smeq 0$), zero energy solutions are Majorana fermions. For the Floquet case ($\xi_1\neq0$), zero or $\hbar \omega/2$ quasienergy states are Floquet Majorana fermions (FMFs).\cite{Cirac2011prl}

We apply the superlattice-based method to the time dependent Schroedinger equation $\left(\hat{H}(t)\smmi \ci\hbar\frac{d}{dt} \right)\ketLR{\Psi(t)}=0$ by following Sec.\ref{SC:Floquet}. After defining a field operator $\Upsilon^\dagger=(\hat{d}^\dagger,\hat{d},\hat{\gamma}_1,\hat{\gamma}_2)$, we rewrite the Hamiltonian as $\hat{H}(t_i)= \frac{1}{2}\Upsilon^\dagger \hat{H}_i \Upsilon$, where $t_i\smeq \Delta t$ (with $2\pi/\omega\smeq T \smeq P \Delta t$) and $\hat{H}_i$ is represented by the matrix
\be
\hat{H}_i=\left(\begin{array}{cccc}
\xi_\mathrm{d}(t_i)&0&-g^*&0\\
0&-\xi_\mathrm{d}(t_i)&g&0\\
-g&g^*&0&-\ci \eta/2\\
0&0&\ci \eta/2&0
\end{array}\right).
\ee
The model consists of a sequence of layers (with four effective sites each) connected by identical hopping matrices ${\hat{T}_{i,i+1}}\smeq \frac{\ci\hbar}{2\Delta t} \eins_4$ originated from to the time derivative in the Schroedinger equation. Given the superlattice structure $\hat{H}_i\smeq \hat{H}_{i+P}$, we compute the characteristic matrix $\mathcal{M}$ at zero artificial energy in order to find the quasienergy states by using Eq.\eqref{EQ:Quasienergies} (see Sec.\ref{SC:Floquet}). In this example, we implement a superlattice with $P\smeq 150$ layers: The criteria in Eq.\eqref{EQ:condition} are fulfilled for the parameters used in Fig.\ref{FG:fig6}.

Without loss of generality, we choose a real $g$ and investigate either cases of dominant $\eta$ (relatively shot wire) and dominant $g$ (relatively strong dot-wire coupling). This are shown in Figs.\ref{FG:fig6}(b-e) and \ref{FG:fig6}(d-g), respectively. We first plot the energies in absence of the driving ($\xi_1\smeq 0$) as a function of $\xi_0$: There, the Majorana condition is satisfied for $\xi_0\smeq 0$. However, such MFs are not robust to perturbations in $\xi_0$ [expected from electric noise in the electrostatic gate $V_g$, see Fig.\ref{FG:fig6}(a)] as the energy dependence of the associated levels is linear in $\xi_0$. As shown in Figs.\ref{FG:fig6}(b) and \ref{FG:fig6}(e), we define $\hbar\omega_c$ as the energy difference between the MFs and the excited states at $\xi_0\smeq 0$, which is given by $\hbar\omega_c \equiv \sqrt{\eta^2/4+2 g^2}$.

In Figs.\ref{FG:fig6}(c-d) and \ref{FG:fig6}(f-g), we plot the quasienergies $\ve$ as a function of $\xi_0$ for driving amplitudes $\xi_1\smeq 0.1 \hbar\omega_c$ and $\xi_1\smeq 0.3 \hbar\omega_c$. The results correspond to driving frequencies $\omega$ falling above, below and at the critical value $\omega_c$. In all cases, FMFs solutions appear either at vanishing and finite $\xi_0$ (including FMFs with quasienergy $\hbar\omega/2$).\cite{Cirac2011prl} In the general case, the FMFs show a sensitivity on $\xi_0$ similar to that shown by static MFs as the relevant $\ve(\xi_0)$ is linear in $\xi_0$.
However, we find that vanishing-$\xi_0$ FMFs become less sensitive to the dot's energy at the critical frequency, i.e., $\omega\smeq\omega_c$. This improves as the driving amplitude $\xi_1$ increases. This result illustrates the superlattice version of the eigenvalue method applied to a problem that combines topological superconductors, quantum dots and Floquet physics.

\section{Conclusions}
\label{SC:Concl}

We revisited the eigenvalue method---originally proposed for the study of electronic states in  superlattices and translationally invariant or Bloch systems---and presented a construction that allows for its application to the study of periodically driven systems. Our scheme, which was applied with success to obtain the results reported in Ref.\onlinecite{ReynosoFrustaglia2013a}, treats the physical time $t$ as an additional spatial dimension, $\tilde{x}$. The physical time is discretized generating a lattice system with an additional dimension. The enlarged system is time-independent along an artificially added time: Its solutions can be classified using the artificial energy, $\tilde{E}$. We showed that the solutions to the original time-dependent problem are directly obtained from the $\tilde{E}\smeq 0$ solutions with positive velocity along $\tilde{x}$. The quasienergies of the original Floquet problem are encoded in the quasimomenta along $\tilde{x}$ for each of these $\tilde{E}\smeq 0$ solutions. We established the criteria (see Eq.\eqref{EQ:condition}) for an accurate description of the Floquet problem with the lattice construction.

We notice that adding an extra dimension in the treatment of Floquet problems is also natural within frequency domain treatments. Insight into Floquet topology can be gained by treating the frequency domain as a spatial dimension as all energy states of the enlarged system are relevant to the Floquet problem.\cite{Sadreev2012,GomezPlatero2013} Here, instead, we followed a different strategy by working in the time domain: In contrast to the frequency domain, our treatment introduces a truly periodical dependence in the extra dimension (treated as spatial) which holds in the full parameter space. It is precisely due to the retained periodicity that we can attack the Floquet problem with methods originally designed for Bloch systems. No claims of greater efficiency are made for this method (demanding similar resources to those needed by finding of Floquet solutions from the eigenvectors of the evolution operator over one driving period). The advantage relies on its accessibility: We showed that existing implementations for solving Bloch superlattices can be readily adapted to solve Floquet problems. The scheme provides the Floquet solutions directly from the eigenstates of the enlarged system at a single artificial energy, this is different from the frequency-based approach in which the full spectrum of the enlarged system is required.\cite{Sambe1973}

We provided several examples of the eigenvalue method. We first applied the method within its original context: time-independent systems. We studied a multichannel quantum wire subject to spin-orbit coupling in the vicinity of a superconductor. By introducing an additional magnetic field along the wire's axis $x$, we discussed the development of topological superconducting phases where Majorana excitations can exist as edge states. In this case, the method was useful to: (a) obtain the dispersion relations, (b) obtain the effective superconducting gap and thereby identifying potential topological transitions, and (c) finding--- by using the evanescent matrix--- the topological number by counting the number of MFs that can exist at the edge of the sample. This allowed us to see that the SOC term proportional to $p_y\sigma_x$ may generate gap closings leading to topological phase transitions. Similarly, by applying the superlattice version of the method, we showed that the presence of a periodic potential along $x$ can break the topological superconducting phase. Finally, as an example of the enlarged lattice construction  that allows to deal with Floquet systems, we studied the effects of microwave excitations on a quantum dot coupled to a topological superconductor finite sample. We showed that there exists a critical frequency at which the Majorana condition becomes less sensitive to the dot's energy. Larger microwave power would reduce such sensitivity even further.

\acknowledgments
We acknowledge useful discussions with A. Doherty and K. Flensberg. AAR acknowledges support from the Australian Research Council Centre of Excellence scheme CE110001013, from ARO/IARPA project W911NF-10-1-0330, and the hospitality of University of Seville. DF acknowledges support from the Ram\'on y Cajal program, from the Spanish Ministry of Science and Innovation's project No. FIS2011-29400, and from the Junta de Andaluc\'ia's Excellence Project No. P07-FQM-3037.

\appendix
\section{Quantum wire Hamiltonians}
\subsection{Normal quantum wire}
\label{AP:TBnormal}

Following standard finite-differences, we proceed by writing the Hamiltonian Eq.\eqref{EQ:Hcont} on a 2D square lattice similar to the one shown in Fig.\ref{FG:fig2}(a). The site coordinates are $(x_i,y_j) = (i a_0, j a_0)$ with $a_0$ the lattice spacing, $j=1,2,..,N_y$, $W/a_0\smeq N_y\smpl 1$ and $i\in \mathbb{Z}$. The $N_y$ spatial sites per layer imply $N\smeq 2 N_y$ effective sites given the spin degree of freedom. For a layer at coordinate $x_i$, we define
\be
t_H\equiv \frac{\hbar^2}{2m^*a_0^2}~,~~
\varsigma_\parallel \equiv \frac{\alpha_\parallel}{2a_0}~,~~\varsigma_\perp \equiv \frac{\alpha_\perp}{2a_0}~,
\ee
and the $2\times2$ matrices
\bea
\hat{H}_i^{y_{j}}\equiv \left(
\begin{array}{cc}
4  t_H+ V(x_i,y_j) & E_Z \\
E_Z & 4  t_H+ V(x_i,y_j)
\end{array} \right)~,\notag  \\\hat{T}_\parallel^{x_i y_{j}}\equiv \left(
\begin{array}{cc}
-t_H & \varsigma_\parallel \\
-\varsigma_\parallel & -t_H
\end{array} \right)~,~~\hat{T}_\perp^{x_i y_{j}}\equiv\left(
\begin{array}{cc}
-t_H & \ci \varsigma_\perp \\
\ci\varsigma_\perp & -t_H
\end{array} \right).
\eea
For generality we have introduced a potential $V(x,y)$ to the Hamiltonian, this allows for an eventual introduction of superlattice potentials. These matrices (operating on spin-$\frac{1}{2}$ spinors in the $z$ basis) are used to assemble the Hamiltonians
\bese
\bea
\hat{H}_i &=& \left(
\begin{array}{ccccc}  \ddots & \vdots  &  \vdots  & \vdots   &  \Ddots \\ \cdots & \hat{H}_i^{y_{j-1}}  & \hat{T}^{x_i y_{j-1}}_\perp & \zeins_{2} & \cdots
\\ \cdots   &  (\hat{T}_\perp^{x_i y_{j-1}})^\dagger & \hat{H}_i^{ y_{j}} & \hat{T}_\perp^{x_i y_{j}} & \cdots
 \\ \cdots &  \zeins_{2} & (\hat{T}_\perp^{x_i y_{j}})^\dagger & \hat{H}_i^{ y_{j+1}} &  \cdots \\ \Ddots & \vdots  &  \vdots  & \vdots   &  \ddots
\end{array} \right),\\
\hat{T}_{i,i+1} &=& \left(
\begin{array}{ccccc}  \ddots & \vdots  &  \vdots  & \vdots   &  \Ddots \\ \cdots & \hat{T}^{x_i y_{j-1}}_\parallel  & \zeins_{2} & \zeins_{2} & \cdots
\\ \cdots   & \zeins_{2}& \hat{T}_\parallel^{x_i y_{j}} & \zeins_{2} & \cdots
 \\ \cdots &  \zeins_{2} & \zeins_{2} &\hat{T}_\parallel^{x_i y_{j+1}}  &  \cdots \\ \Ddots & \vdots  &  \vdots  & \vdots   &  \ddots
\end{array} \right).
\eea
\label{EQ:APconst}
\eese
These are $2 N_y$-dimensional square matrices. The vector encoding the solution at the layer located at $x_i$ (see Eq.\eqref{EQ:Hfull}) is $G_{i}=(g_{i,1},g_{i,2}, \cdots,g_{i,N})^T$ with $(g_{i,2(j-1)+1},g_{i,2(j-1)+2})^T=(\psi_\uparrow(x_i,y_j),\psi_\downarrow(x_i,y_j))^T$, i.e., the spin-$\frac{1}{2}$ wavefunction at site $(x_i,y_j)$. Here we are considering only position independent SOC so that the superscript in $\hat{T}_\parallel^{x_i y_{j}}$ and $\hat{T}_\perp^{x_i y_{j}}$ turns irrelevant (though kept for generality). Furthermore, for vanishing $V(x,y)$ (as presented in Fig.\ref{FG:fig2}) there is no dependence on $x_i$, i.e., $P=1$ and the characteristic matrix $\mathcal{M}(E)$ of Eq.\eqref{EQ:TransMat} is obtained by taking $\hat{H}_0\smeq \hat{H}_i$ and $\hat{T}_0\smeq \hat{T}_{i,i+1}$.

\subsection{Superconducting quantum wire}
\label{AP:TBsup}

The BdG equation describing the mean field $s$-wave superconducting pairing, when projected on the lattice representation, leads to the Hamiltonian\cite{Cuevas96}
\be
\hat{H}_\Delta=-\mu \sum_{i,j,\sigma}\hat{c}_{i,j,\sigma}^\dagger \hat{c}_{i,j,\sigma} + \Delta \sum_{i,j}\left(  \hat{c}_{i,j,\downarrow} \hat{c}_{i,j,\uparrow}+ \hat{c}_{i,j,\uparrow}^\dagger \hat{c}_{i,j,\downarrow}^\dagger\right),
\ee
with $\Delta=\Delta_0$ assumed real and $\hat{c}_{i,j,\sigma}$ the electronic annihilation operator at site $(x_i,x_j)$ with spin projection $\sigma$ along the $z$ axis. The $\hat{H}_\Delta$ must be added to the normal Hamiltonian $\hat{H}_N$, coupling the electron block to the hole block, since
\be
\left[\hat{c}_{i,j,\downarrow},\hat{H}_\Delta+\hat{H}_N\right]=\left[\hat{c}_{i,j,\downarrow},\hat{H}_N\right]-\mu \hat{c}_{i,j,\downarrow} + \Delta \hat{c}_{i,j,\uparrow}^\dagger~.
\ee
We define a 4-dimensional spinor with electron ($\psi$) and hole ($\psi^\dagger$) sectors: The wavefunction at position $x_i,x_j$ is then written as $(g_{i,4(j-1)+1},g_{i,4(j-1)+2},g_{i,4(j-1)+3},g_{i,4(j-1)+4})^T=(\psi_\uparrow(x_i,y_j),\psi_\downarrow(x_i,y_j),\psi^\dagger_\downarrow(x_i,y_j),-\psi^\dagger_\uparrow(x_i,y_j))^T$.
Notice that even when $j=1,2,..,N_y$ (as in the normal case), the number of effective sites per layer is doubled to $N=4N_y$ and the wavefunction at position $x_i$ becomes $G_{i}=(g_{i,1},g_{i,2}, \cdots,g_{i,4N_y})^T$. By using this basis, we define the $4\times4$ matrices
\begin{widetext}
\bea
\hat{H}_i^{y_{j}}\equiv \left(
\begin{array}{cccc}
4  t_H+ V(x_i,y_j) -\mu& E_Z & \Delta_0 & 0 \\
E_Z & 4  t_H+ V(x_i,y_j)-\mu & 0 & \Delta_0  \\
\Delta_0 & 0 & \mu- 4  t_H- V(x_i,y_j) & E_Z \\
0 & \Delta_0 & E_Z &\mu- 4  t_H- V(x_i,y_j)
\end{array} \right)~,\notag  \\\hat{T}_\parallel^{x_i y_{j}}\equiv \left(
\begin{array}{cccc}
-t_H & \varsigma_\parallel &0&0\\
-\varsigma_\parallel & -t_H&0&0\\
0&0&t_H & -\varsigma_\parallel \\
0&0&\varsigma_\parallel & t_H
\end{array} \right)~,~~\hat{T}_\perp^{x_i y_{j}}\equiv\left(
\begin{array}{cccc}
-t_H & \ci \varsigma_\perp &0&0\\
\ci\varsigma_\perp & -t_H&0&0 \\
0&0& t_H & -\ci \varsigma_\perp \\
0&0& -\ci \varsigma_\perp & t_H
\end{array} \right).
\eea
With these matrices, we proceed to build the $4N_y \times 4N_y$ layer Hamiltonian $\hat{H}_i$ and hopping matrices $\hat{T}_{i,i+1}$. The structure is similar to that of Eq.\eqref{EQ:APconst} but based on $4\times4$ blocks, instead, after replacing the $~\zeins_2$ with $~\zeins_4$. As in the normal case, $P=1$ for vanishing $V(x,y)$ and the characteristic matrix can be obtained from any $\hat{T}_{i,i+1}$ and $\hat{H}_i$ as they do not depend on $x_i$.

Once an eigenstate is obtained, we compute the electron-hole polarization, $P_{eh}$, summing over all the $y_j$ for a given $x_i$ as
\be
P_{eh}=\frac{A_e-A_h}{A_e+A_h}~,~~
A_e=\sum_{j=1}^{N_y} \left( \left|g_{i,4(j-1)+1}\right|^2+\left|g_{i,4(j-1)+2}\right|^2 \right)~,~~A_h=\sum_{j=1}^{N_y} \left( \left|g_{i,4(j-1)+3}\right|^2+\left|g_{i,4(j-1)+4}\right|^2 \right),
\ee
where $A_h$ and $A_e$ are the electron and hole contribution, respectively.
\end{widetext}


%

\end{document}